\documentclass[9pt,twocolumn,twoside,lineno]{pnas-new}

\templatetype{pnasresearcharticle} 

\title{Time-evolving controllability of effective connectivity networks during seizure progression}

\author[a,b]{Brittany H. Scheid}
\author[a,b]{Arian Ashourvan} 
\author[a,c]{Jennifer Stiso} 
\author[a,b,d]{Kathryn A. Davis}
\author[a,b,d]{Fadi Mikhail} 
\author[e]{Fabio Pasqualetti}
\author[a,b,d]{Brian Litt} 
\author[a,b,f,g,h,i]{Danielle S. Bassett}

\affil[a]{Department of Bioengineering, University of Pennsylvania, Philadelphia, PA 19104}
\affil[b]{Center for Neuroengineering and Therapeutics, University of Pennsylvania, Philadelphia, PA 19104}
\affil[c]{Department of Neuroscience, University of Pennsylvania, Philadelphia, PA 19104}
\affil[d]{Department of Neurology, Hospital of the University of Pennsylvania, Philadelphia, PA 19104}
\affil[e]{Department of Mechanical Engineering, University of California, Riverside, CA 92521}
\affil[f]{Department of Electrical \& Systems Engineering, Philadelphia, PA 19104}
\affil[g]{Department of Physics \& Astronomy, University of Pennsylvania, Philadelphia, PA 19104}
\affil[h]{Department of Psychiatry, Perelman School of Medicine, Philadelphia, PA, 19104}
\affil[i]{Santa Fe Institute, Santa Fe, NM 87501}
\leadauthor{Scheid} 

\begin{abstract}
Over one third of the estimated 3 million people with epilepsy in the US are medication resistant. Responsive neurostimulation from chronically implanted electrodes provides a promising treatment option and alternative to resective surgery. However, determining personalized optimal stimulation parameters, including when and where to intervene to guarantee a positive patient outcome, is a major open challenge. Network neuroscience and control theory offer useful tools that may guide improvements in parameter selection for control of anomalous neural activity. Here we use a novel method to characterize dynamic controllability across consecutive effective connectivity (EC) networks based on regularized partial correlations between implanted electrodes during the onset, propagation, and termination phases of thirty-four seizures. We estimate regularized partial correlation adjacency matrices from one-second time windows of intracranial electrocorticography recordings using the Graphical Least Absolute Shrinkage and Selection Operator (GLASSO). Average and modal controllability metrics calculated from each resulting EC network track the time-varying controllability of the brain on an evolving landscape of conditionally dependent network interactions. We show that average controllability increases throughout a seizure and is negatively correlated with modal controllability throughout. Furthermore, our results support the hypothesis that the energy required to drive the brain to a seizure-free state from an ictal state is smallest during seizure onset; yet, we find that applying control energy at electrodes in the seizure onset zone may not always be energetically favorable. Our work suggests that a low-complexity model of time-evolving controllability may offer new insights for developing and improving control strategies targeting seizure suppression.                           
\end{abstract}

\dates{This manuscript was compiled on \today}

\begin{document}

\maketitle
\thispagestyle{firststyle}
\ifthenelse{\boolean{shortarticle}}{\ifthenelse{\boolean{singlecolumn}}{\abscontentformatted}{\abscontent}}{}

\dropcap {R}esponsive neurostimulation (RNS) is a non-resective treatment for medication-resistant epilepsy that aims to suppress seizures by delivering an electrical stimulus through intracranial electrodes in response to abnormal electrographic activity. Despite the fact that the first implantable RNS device was approved in 2013, the mechanism of action and optimal patient-specific stimulation settings remain unknown \cite{Bergey2015, Hartshorn2018}. A better understanding of the cortical locations and patterns of brain activity that are most responsive to stimulation would result in increased treatment efficacy and efficiency. By modelling the brain as a network, with distinct brain regions representing nodes and with statistical dependencies representing measures of influence between regions as edges, the challenge of finding the best strategy for seizure suppression can be recast as a problem of network control \cite{Pasqualetti2014,Tang2019}. 

Evidence from prior empirical and computational studies suggests that a linear network control framework built on the pattern of white matter tract connections in large-scale cortical and subcortical areas can be used to predict the response of brain activity to an external stimulus \cite{Muldoon2016, Khambhati2018, Stiso2018, Kim2018}. The theory also provides predictions of the optimal stimulation parameters required to drive activity from an initial brain state to a desired target state if the network is controllable \cite{Gu2015,Betzel2016}. These predictions have been quantified for direct electrical stimulation targeted at improving memory \cite{Stiso2018}, but have not yet been defined or tested for stimulation applied to controlling seizures for three main reasons. First, seizures are characterized by rapid state transitions that violate the assumption of linearity over the course of the whole seizure \cite{Smith2011, Kramer2012}. Second, while controllability is defined for both positive and negative edge weights, it is typically modeled on positively-weighted networks derived from structural imaging techniques that do not characterize the inhibitory pathways between brain regions, which may play an essential role in seizure termination \cite{Timofeev2004,Loddenkemper2014}. Finally, it has been theorized and observed that information propagation through the brain may depend on the underlying state of synchrony between brain regions, and that exogenous inputs to brain regions may bias the direction of information flow \cite{Kirst2016, Palmigiano2017}. In light of these data, it is clear that informing estimates of controllability with static white matter networks fails to capture dynamic interareal influences over changing patterns of network coherence \cite{Battaglia2012}.  

\begin{figure}[t]
\centering
\includegraphics[width=\linewidth]{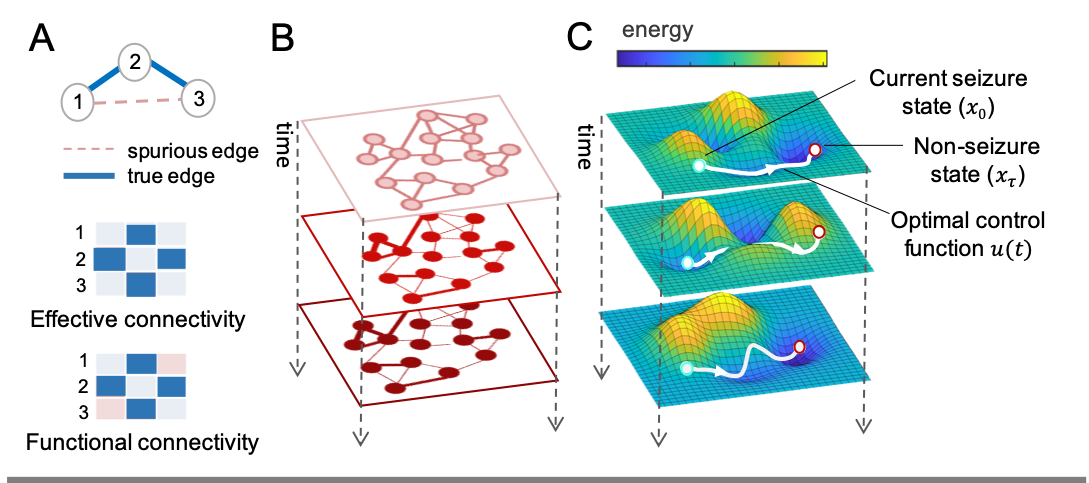}
\caption{\textbf{Time evolving effective connectivity as a tool in seizure control strategies}. (A) In a toy network with two empirical functional connections across three nodes, correlative functional connectivity may estimate a spurious edge between nodes that interact indirectly. Conversely, effective connectivity (EC) reflects the \emph{direct influence} between nodes in a network, setting the edge weight of conditionally independent nodes to zero, and thus provides a scaffold to model control strategies over time. Throughout a seizure, we can build EC networks to represent distinct phases of neural connectivity (B), and infer a dynamic control-energy landscape from those networks (C). We can then use tools from control theory to determine the optimal control input required to drive neural activity to seizure freedom given the state of the control energy landscape.} \label{fig:soexciting}
\end{figure}

We address the above limitations in a novel approach that measures dynamic controllability in effective connectivity (EC) networks built from electrographic recordings in sequential time windows. Connectivity measures derived from brain activity fall on a spectrum from functional connectivity, which does not distinguish between direct vs. indirect node correlations, to effective connectivity, in which the connectivity structure implies underlying causality \cite{Horwitz2003}. A number of data-driven methods including Granger causality, directed transfer function, and partial directed coherence have been used to create EC brain networks \cite{Sakkalis2011,Ren2019}. In the context of this paper, effective connectivity describes the conditional dependence of nodes in a network, given the activity of all other nodes, and is therefore a powerful approach for understanding the dynamics of seizure spread throughout an epileptogenic network \cite{Sakkalis2011} (Fig. \ref{fig:soexciting}). Specifically, we use a penalized form of the standardized inverse covariance as an estimate of undirected conditional dependence between network nodes, as it has been shown to estimate the presence of true network connections with high sensitivity compared to other methods \cite{Smith2011, Friedman2008, Rosa2015, Das2019}. 

Here we build on prior work in network control theory to address several specific hypotheses regarding seizure dynamics in humans. We focus on two measures of controllability that characterize the ease of driving the brain into new states as time evolves \cite{Pasqualetti2014,Karrer2019}. \textit{Average controllability} reflects the average energy input required at a node to transition the activity from some initial brain state to all other possible final states \cite{Gu2015,shine2019human}. \textit{Modal controllability} reflects the ease of moving a system from a starting state to a difficult-to-reach state \cite{Gu2015, Wu-Yan2018, Khambhati2018}, and quantifies how the spectral properties of the underlying network will distribute input energy throughout the system. Modal controllability can be separated into the measures of \textit{persistent modal controllability} and \textit{transient modal controllability}, which describe whether energy input at a node is likely to perturb the slow- or fast-damping modes of the system, respectively \cite{Tang2019, Karrer2019, Stiso2018}. Collectively, these statistics allow a broad assessment of a system's accessible control strategies. We hypothesize that seizure propagation will be accompanied by a heightened average controllability and decremented modal controllability consistent with an enhanced strengthening of the epileptic network \cite{Gu2015, Wu-Yan2018}. We extend this evaluation by using a model of seizure-phase-dependent linear control to find the optimal control function required to drive the brain to a seizure-free state from a subset of control nodes. We predict that controlling the nodes in the seizure onset zone (SOZ) as epileptiform activity emerges will be the most effective. Our prediction is built upon prior observations that these nodes precede the rest of the network in showing increased functional connectivity \cite{Khambhati2016}, and we reasoned that such precedence may facilitate their local control before the seizure spreads. Broadly, our study offers a principled approach to the study of time-evolving coherence-mediated control, which has the potential to inform practical efforts to improve neurostimulation-based therapies.

\section*{Results}
\subsection*{Characterization of Seizure Phases} 
\begin{figure}[bt]
\centering
\includegraphics[width=\linewidth]{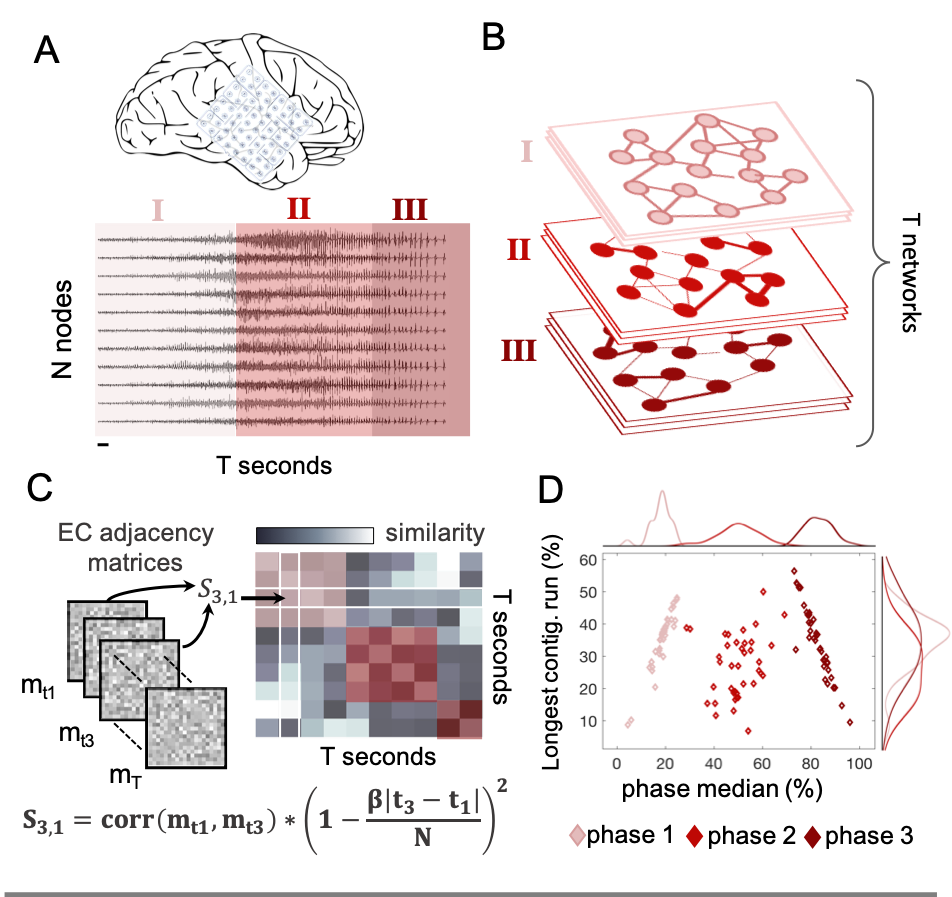}
\caption{\textbf{Time evolving controllability through effective connectivity}. \emph{(A)} For each seizure, we extracted consecutive 1-sec time windows of cortical ECoG recordings from $N$ electrode channels. \emph{(B)} From these data, we estimated $T$ effective connectivity networks. \emph{(C)} We used community detection to determine three seizure phases based on the similarity of the regularized partial correlation adjacency matrices. \emph{(D)} The distribution of the three largest phases found in all 34 seizures. Phases are organized chronologically and are plotted by their normalized temporal median versus the longest consecutive run of time windows, as a percentage of total seizure length. Each data point represents a phase in a single seizure.}
\label{fig:methods}
\end{figure}

Prior studies of \textit{functional} connectivity networks built with ECoG data suggest that seizures can be segmented into three main phases: onset, propagation, and termination \cite{Khambhati2015, Kramer2012, Jiruska2013}. We sought to determine whether such a phase separation existed in EC; if so, dynamic controllability could be meaningfully compared across seizures by segregating each subject's data according to seizure phase. To address this question, we began by constructing EC networks from consecutive 1-sec time windows of preictal and ictal ECoG recordings from 34 partial or secondarily generalized seizures across 14 subjects undergoing EEG monitoring prior to surgical treatment for medication-resistant epilepsy. EC networks were obtained using the GLASSO method to estimate the regularized partial correlation of the recorded timeseries in each time window (see Methods for more details). For each seizure of variable length $T$ seconds recorded with a variable $N$ channels, this procedure provided a total of $T$ regularized partial correlation matrices of size $N \times N$ (see Fig. \ref{fig:methods}A, B, and Methods). 

To detect seizure phases, we next calculated a $T \times T$ similarity matrix, encoding a weighted network, where the $ij$-th element represented the Pearson correlation coefficient between the upper triangles of the $i$-th and $j$-th EC networks, inversely weighted by a constant $\beta$ times the duration between timepoints to promote state contiguity (see Methods). Using a Louvain-like locally greedy algorithm to maximize a modularity quality function, we partitioned the columns of the similarity matrix thereby separating each seizure into three temporal phases characterized by distinguishable patterns of EC (Fig. \ref{fig:methods}C). The largest temporally contiguous network assignments in each community were selected for the following analyses, and were labeled chronologically as seizure phases 1 through 3 with respect to the median time point of the community cluster (Fig. \ref{fig:methods}D). On average, 96 ± 5\% of the EC networks from a given seizure were assigned to one of the three contiguous phases, compared with only 30 ± 12\% assigned during an equal-length preictal period where distinct phases would not be expected (Fig. S2). Thus we found that our data-driven method could be used to demarcate onset, generalization, and termination phases across seizures, and that the pattern of region-to-region influences were sustained within each seizure phase.   

\begin{figure*}[!t]
\centering
\includegraphics[width=\linewidth]{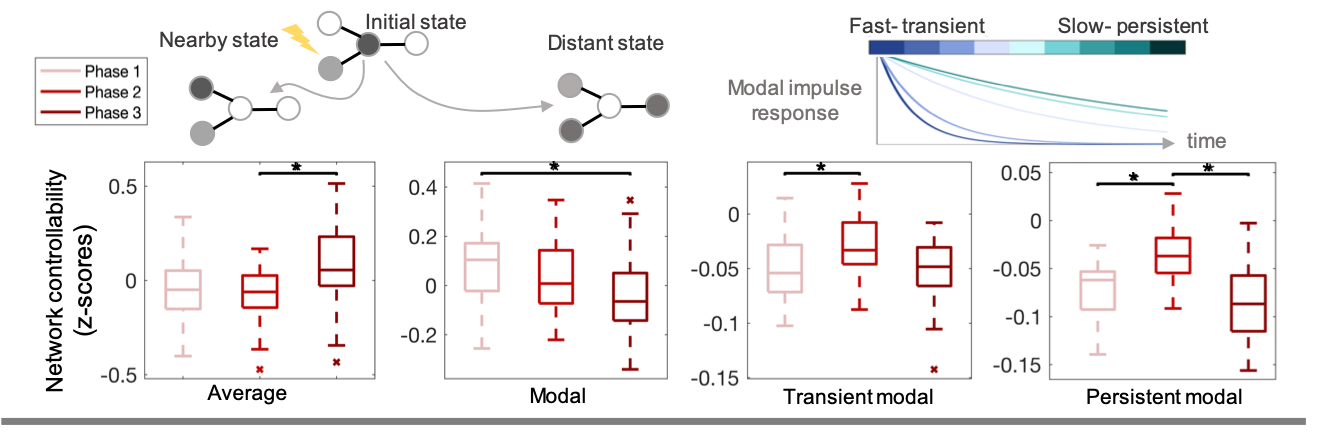}
\caption{\textbf{Group-level controllability dynamics throughout early, middle, and late seizure phases.} 
A single controllability metric value was obtained for each phase in a seizure, and the distribution of phase metric values across all seizures ($N=34$) is shown. Average controllability indicates the ease of driving network activity to nearby states, and was found to increase throughout seizure phases $({\chi}^2(2, 34)=9.59,\ p<0.01)$. Modal controllability indicates the ease of driving network activity to hard-to-reach, or distant states, and was found to decrease throughout seizure phases $({\chi}^2(2, 34)=9.59,\ p<0.01)$. Transient and persistent modal controllability describe the ease of perturbing the slow, sustaining modes of the system, or the fast, attenuating modes. A significant effect of seizure phase was found for both transient modal controllability $({\chi}^2(2, 34)=8.88,\ p<0.016)$, and persistent modal controllability values $({\chi}^2(2, 34)=18.59,\ p<1\times10^{-4})$. Boxplots indicate the 75\% confidence interval(box), median (solid line), 95\% confidence interval (whiskers), and outliers (stars). Starred bars indicate significant metric differences between phases at the $p<0.017$ level, determined using Friedman's ANOVA.}
\label{fig:control_graphs}
\end{figure*}

\subsection*{Dynamic Controllability of Epileptic Networks} We now turn to an investigation of how the effective connectivity network representing each seizure phase impacts phase-dependent controllability. In the context of our model, controllability analysis can answer the question of whether energy input during a given phase will be characterized by diffuse propagation throughout the network, pushing the brain state into other nearby states (high average controllability), or whether input energy is more likely to be channeled towards brain states that are hard to reach and require changes to the dynamic modes of the system for control (high modal controllability). Additional analysis of persistent modal and transient modal controllability metrics can describe the extent to which input energy into the brain network is either sustained by affecting slowly decaying modes, or attenuated by affecting quickly decaying modes, respectively.

\par For all controllability statistics, we observed a significant phase dependence at the group level. We calculated average, modal, transient modal, and persistent modal controllability metrics for each brain region in each EC network, resulting in an $N \times 1$ vector per calculation. The vectors for a given metric were assigned to one of the three seizure phases, and we calculated the median controllability values across nodes in each phase for each seizure to arrive at representative nodal phase values. These nodal values were then averaged to give a single network value in each seizure phase. Across 34 seizures in 14 subjects with epilepsy, we observed a significant effect of seizure phase on average controllability $({\chi}^2(2, 34)=9.59,\ p<0.01)$, modal controllability $({\chi}^2(2, 34)=9.59,\ p<0.01)$, transient modal controllability $({\chi}^2(2, 34)=8.88,\ p<0.016)$, and persistent modal controllability ${(\chi}^2(2, 34)=18.59,\ p<1\times10^{-4})$ using Friedman’s analysis of variance (ANOVA) across phases. In \emph{post-hoc} analysis after Bonferroni correction for multiple comparisons of seizure phases within each metric, we found a significant increase in average controllability from propagation to termination phases ($N=34,\ t=-3.03,\ p<0.01$), with a steady and significant decrease in modal controllability from onset to termination ($N=34,\ t=3.03,\ p<0.01$). Both transient and persistent modal controllability demonstrated a significant increase in value from onset to propagation ($N=34,\ t=-2.78,\ p<0.016$, and $t=-3.15,\ p<0.01$, respectively), and both decreased again from propagation to termination phases ($t=2.3,\ p< 0.064$, and $t=-4.12,\ p<0.001$) though the decrease in transient controllability did not reach statistical significance (Fig. \ref{fig:control_graphs}). Our results demonstrate that the extent of input energy propagation throughout the network will increase with phase progression, and that ease of directing activity towards hard-to-reach brain states is greatest at seizure onset. 


\subsection*{Phase-Dependent Optimal Control Energy} The linear dynamical model used to understand state-invariant controllability in brain networks can also be used to identify the optimal control energy necessary to drive the brain from an initial \textit{brain state} $x_0$, to a final state $x_{\tau}$ over ${\tau}$ time steps \cite{Gu2015,Stiso2018,Betzel2016,Tang2019}. In contrast to the \textit{seizure phase}, which is defined by a particular pattern of interareal influence reflected in the structure of the EC networks, the \textit{brain state} can describe the particular signals or information patterns that move across these networks. We used the optimal control framework to ask \textit{when} during the seizure it would be most energetically favorable to externally drive the brain activity to a seizure-free baseline. Given that current RNS treatment practices have demonstrated clinical success when injecting current just after seizure onset is detected, we hypothesized that control energy would be smallest in the seizure onset phase when the seizure is often spatially confined \cite{Morrell2011}.  

\par
\begin{figure*}[bthp]
\centering
\includegraphics[width=\linewidth]{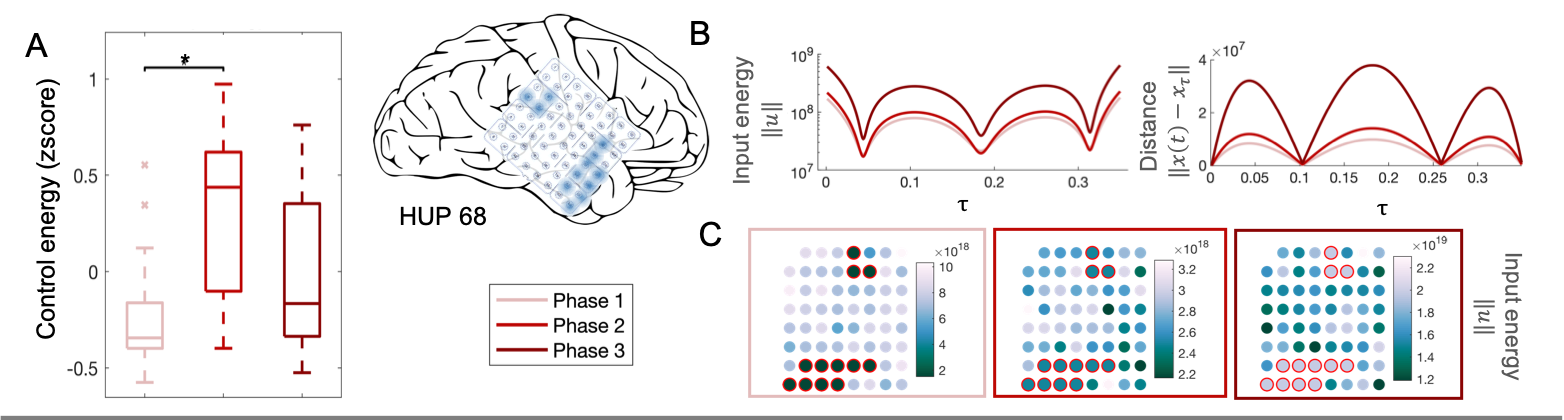}
\caption{\textbf{State-dependent optimal control energy.} \emph{(A)} A significant effect of optimal control energy on seizure phase was found at the group-level$({\chi}^2(2, 34)=18.41,\ p<1\times10^{-4})$; energy values were significantly lower in the onset phase compared to seizure propagation. Boxplots indicate the 75\% confidence interval (box), median (solid line), 95\% confidence interval (whiskers), and outliers (stars). Starred bars indicate significant metric differences between phases using a t-test at the $p<0.017$ level. \emph{(B)} The trajectory from seizure state to preictal baseline is controlled through designated SOZ nodes highlighted in blue on the cortical grid for a seizure in subject HUP68 over ${\tau}$ control time steps. Left: Total input energy into the set of SOZ nodes for subject HUP68 across ${\tau}$ control time steps. Right: The Euclidean distance of network state at time $t$ to final target network state $x_{\tau}$ along the control trajectory (estimation error: $31.46$). \emph{(C)} Optimal energy values are displayed across the three phases for a single seizure in subject HUP68, where SOZ energy was significantly lower than the null distribution in the onset phase ($N=5000,\ p<0.05$). Nodes are spatially arranged according to physical electrode placement, and the color of each node outside the SOZ reflects the optimal control energy value averaged across resampled control sets in which the node participated. Nodes in the SOZ are outlined in red.
 SOZ\-- seizure onset zone}
\label{fig:optimal_control}
\end{figure*}

 We measured the optimal control energy required for a single node to move the brain state recorded during seizure onset, propagation, or termination, respectively, to the baseline brain activity level recorded in the preictal period. We defined brain state using high-$\gamma$ bandpower (30-150Hz) of the ECoG channel recordings averaged over each one-second time window in a phase, as it has been shown to capture population spiking frequencies at the seizure core \cite{Weiss2015} and is sensitive to variation of both signal frequency and amplitude \cite{Weiss2013, Baldassano2017, Stiso2018}. We averaged the optimal energy across all nodes in a particular phase, and then compared this average value across the same phase in all other seizures (Fig. \ref{fig:optimal_control}A). Again Friedman’s ANOVA was conducted to find that seizure phase had a significant effect on optimal control energy $({\chi}^2(2, 34)=18.41,\ p<1\times10^{-4})$. Specifically, the seizure onset phase showed a significantly lower control energy requirement than the propagation phase at the group level $(N=34,\ t=-4.24,\ p<6.6\times10^{-5})$, and a lower energy than in the termination phase, although not significantly so $(N=34,\ t=2.67,\ p<0.02)$. The results demonstrate how a perspective of dynamic controllability can identify seizure phases that may be controlled in an energetically favorable manner.

\subsection*{Control Localization in Onset Phase}
The optimal control energy results demonstrated that the seizure onset phase is preferred for its low energy requirements. In an effort to find the best control location, we next asked whether controlling nodes at the SOZ required less energy than controlling any other subset of nodes, given that RNS treatments are typically localized to the SOZ \cite{Bergey2015, Morrell2011}. Seizures in all but four subjects exhibited focal onsets, for a total of 22 seizures localized within the area of the cortical grid electrodes and used in this analysis. SOZ electrodes were determined per standard clinical protocol by an epileptologist, and were then reviewed and validated in an intracranial EEG conference attended by the University of Pennsylvania epilepsy faculty. We measured the optimal control energy and brain state trajectory from $x_0$ at a given seizure phase, to a final state $x_\tau$ resulting from simultaneous control at each SOZ node. We then repeated this measurement in 5000 random reassignments of the SOZ labels among the non-SOZ electrodes. We found that in only half of the subjects exhibiting localizable SOZs, control using the SOZ nodes required significantly less energy than control from other subsets of nodes in the seizure onset phase $(N=5000, p<0.05)$ (Fig. \ref{fig:optimal_control}B and Fig. S9). We observed no correlation between SOZ significance and the number of SOZ nodes (Mann-Whitney $U=84.5, p >0.62$), suggesting that our results are not simply due to differences in SOZ size. As an example, we show the spatial distribution of control energy throughout the 8$\times$8 electrode grid array in subject HUP68 where the SOZ energy was significantly lower during seizure onset (Fig. \ref{fig:optimal_control}C). Energy for nodes outside of the SOZ was calculated as the average energy across all permutation tests in which they participated. These results demonstrate that dynamic controllability analysis can answer spatiotemporal questions regarding stimulation for seizure suppression.

\subsection*{Robustness of Results to Selected Parameters } 
Our analysis relies on a number of assumptions, as well as parameterized methods for EC network generation and phase delineation, and a parameterized optimal control energy model. We sought to increase the reliability of our findings by assessing the sensitivity of our results to our assumptions and parameter values. First, given that spatiotemporal patterns of seizures propagating from the same SOZ may be diverse \cite{Kramer2010, Proix2018}, we chose to treat each seizure as an independent data point. We completed an alternative analysis after grouping seizures within subjects and found that the group-level trends of our main findings were preserved, but only persistent and transient modal controllability showed significant differences across phases (Fig. S3). We also varied the sparsity tuning hyperparameter, $\lambda$, used in EC estimation (Fig. S5) and adjusted the emphasis on temporal contiguity of community assignments, $\beta$, when determining phases (Fig. S6). In these analyses we found no significant change in controllability values for any metric after varying each parameter in a neighborhood around the parameter value used in our main results. Finally we reproduced our group-level optimal control energy results using three additional pairs of time horizon ($\tau$) and distance-energy trade-off ($\alpha$) parameters. We found that optimal control estimation error increased but that group-level trends were preserved for decreasing values of $\tau$ (Fig. S8). Taken together, we find that our main results are robust to reasonable parameter variation.  

\section*{Discussion}
For patients with medication-resistant epilepsy, responsive neurostimulation (RNS) as a means of seizure suppression is a relatively new option that provides an alternative to more permanent and invasive choices such as cortical resection or laser ablation. While the majority of patients improve over the course of months to years with treatment, the current trial and error routine for selecting stimulation parameters is far from ideal and only rarely do patients achieve seizure freedom \cite{Bergey2015, Morrell2011, Hartshorn2018}. We completed this study to introduce a new control-theoretic framework for dynamic controllability that could provide support for clinical treatment decisions and highlight the regions and timepoints for stimulation that require the least input.

\subsection*{Emergence of Synchronous Seizure Phases} 
In a challenge to a static brain controllability perspective, a number of studies find that multiple patterns of corticocortical coupling may appear on the same structural foundation, and that time-evolving EC networks may provide a more accurate view of the connectivity landscape \cite{Battaglia2012, Palmigiano2017, Proix2018, Das2019, Shah2019}. Correlative \emph{functional} connectivity models may inadvertently incorporate common inputs to cortical brain regions, such as neural signals from projections of cholinergic and GABA-ergic ascending fibers, that do not reflect direct corticocortical influences \cite{Kandel2000}. By estimating edges in our EC networks as regularized partial correlations between brain regions, we effectively remove contributions of such latent inputs in our analysis. Using this framework, we uncovered a dynamic profile of controllability with implications for the input energy distribution across brain regions in a given seizure phase. Our results demonstrate increasing average controllability and a corresponding decrease in modal controllability throughout the majority of seizures. Low average controllability at onset implies that a widespread, diffuse distribution of input energy is hindered at seizure initiation, complementing findings in functional connectivity that show marked desynchronization at onset due to loss of inhibitory restraint \cite{Kramer2012, Jiruska2013}. Later throughout propagation and into termination phases where synchrony has been shown to increase \cite{Kramer2012, Jiruska2013, Khambhati2016}, our results suggest that it will become harder to direct the brain into energetically unfavorable states (low modal controllability). 

Interestingly, the temporal trend of persistent and transient modal controllability differs from overall modal control in the onset phase. The relatively low metric values at onset reflect the combination of two factors— first that input attenuation governed by the extreme eigenmodes could be relatively slower in general at seizure onset than during propagation, and second that external perturbations may be less aligned with the extreme eigenmodes at onset. Our findings are consistent with a prior investigation of ictal eigenmode stability in a dynamical linear framework showing that relatively faster input attenuation in the slowest eigenmodes occurred during mid-seizure, perhaps reflecting greater synchrony with compensatory inhibition during the propagation period \cite{Mullen2012}. Our results indicate that decoupling between brain regions at onset could make it particularly difficult to control the very fast or very slow dynamics associated with the extreme eigenmodes of the brain's activity during that phase.

\subsection*{Spaciotemporal Control as a Tool for RNS Intervention}
From the perspective of clinical RNS treatment, our model supports the current practice of intervening at seizure onset \cite{Morrell2011}, with group-level results indicating that the amount of control energy required for a transition to a seizure-free state is smallest at onset. Minimizing energy input is an important clinical consideration, as it not only reduces patient exposure to unnecessary stimulation, but may also extend implant lifetime before an invasive procedure is required for battery replacement. Our analysis on locally differentiated optimal control energy between SOZ and non-SOZ nodes found that the SOZ required significantly less energy for seizure suppression in seizures in only half of the subjects. This finding does not support the SOZ as a universally energetically favorable stimulation target, and begs the question of whether energetically favorable points outside the SOZ could still be clinically relevant stimulation targets. Epilepsy is theorized to be a network disorder \cite{Khambhati2015, Lehnertz2016, Spencer2002}; thus it is likely that multiple cortical locations may be targeted to modulate the same pathologic network. This fact combined with evidence suggesting that seizure precursors may appear outside of the SOZ \cite{Lehnertz2016}, that tissue damage at the SOZ may have remote effects on the functionality of other brain regions \cite{Ung2017}, and that vicinity of stimulus location to SOZ is not always significant for patient outcomes \cite{Seeck2013}, indicate that energetically favorable control points selected by our model may still be effective even if outside the SOZ. 

Although current RNS technology is still limited to feature-based detection of abnormal electrographic signals \cite{Hartshorn2018}, our methodology could enrich a number of model-based control approaches that could be adapted to improve seizure suppression in future implantable devices. For example, model predictive control (MPC) is an iterative multivariable control algorithm that alternately estimates the optimal control trajectory between an initial and final state, then applies only a few time steps of the control function $u(t)$, before making another trajectory estimate \cite{Morari1988}. MPC relies on an internal representation of the dynamical environment, and our partial correlation EC model may improve upon prior estimates by reflecting true corticocortical influences. Other recent work has theoretically demonstrated that seizures may be controlled via static output feedback of a linear dynamical system, wherein the underlying model is derived using multivariate autoregression (MVAR) \cite{Ashourvan2020}. This work could be extended by removing spurious links in an MVAR model using a partial correlation constraint that assesses the correlation between a pair of nodes after removing the linear effects of other variables in the same and previous time points \cite{Wild2010}.  Exciting future directions include testing the hypotheses generated from these models using intracranial data from patients with RNS implants, and prospectively in new stimulation paradigms.

\subsection*{Methodological Considerations}
Our model of dynamic controllability has a number of limitations. First, the GLASSO method used to generate EC matrices relies on the input of a positive definite (PD) covariance matrix derived from the windowed ECoG timeseries data. Due to the dimensionality of our data, we used a pairwise covariance estimation, which often led to a non-PD result. In these cases a “nearby” PD covariance structure was substituted for the EC estimate, introducing a potential source of error \cite{Epskamp2018}. Second, we chose to operationalize brain state here as high-$\gamma$ band power. While this measure is commonly used in epilepsy literature, it does not perfectly capture other relevant features of brain activity including broadband power, the slope of the power spectral density, non-sinusoidal features of the LFP, etc. \cite{Colombo2019, Schaworonkow2019}. Additionally, our model uses a single brain state as an estimate of seizure-free brain activity, when in reality it could be represented by many different states. Lastly, our stipulated model of dynamics assumes noise-free, linear network dynamics within each one-second time window. Although the brain is not a linear system, linear approximations on a short timescale can capture broad dynamics in epilepsy while allowing for the application of linear control theory and ease of interpretability \cite{Smith2011, Kramer2012}. While beyond the scope of this work, future exploration into how brain state definition impacts control properties would complement our analyses. Extending this work to models of non-linear control theory could similarly investigate ways to control the brain to a seizure free manifold, rather than a single state, and would complement the hypotheses generated here \cite{Zanudo2017, Wu-Yan2018}.

In conclusion, we employed a novel method of measuring controllability of effective networks and optimal seizure control energy across three temporal seizure stages. Our results present preliminary work in building theoretically informed hypotheses for future empirical validation.

\matmethods{\subsection*{Subject Data} The ECoG data used for this study was recorded from subdural electrodes implanted in 14 epilepsy patients undergoing EEG monitoring prior to surgical treatment for medically refractory epilepsy at the Hospital of the University of Pennsylvania (512 Hz sample rate) or at the Mayo Clinic in Rochester, MN  (500 Hz sampling rate). All subjects included in this study gave written informed consent in accord with the University of Pennsylvania Institutional Review Board and Mayo Clinic Institutional Review Board for inclusion in this study. The implanted surface electrodes (ad Tech Medical Instruments, Racine, WI) were comprised of both grid and strip arrays with 2.3 mm diameter contacts at 10 mm inter-contact spacing. Electrode placement was planned according to clinical protocol by a team of epileptologists with a separate reference electrode placed distant to the site of seizure onset, as in prior studies \cite{Shah2019, Khambhati2015}. For each of the 34 epileptic events analyzed in this study, the following annotations or characterizations were determined as per standard clinical protocol by an epileptologist and then reviewed and validated in an intracranial EEG conference attended by neurology clinicians at the University of Pennsylvania: seizure categorization as complex partial and/or secondarily generalized, the earliest electrographic onset, unequivocal electrographic onset, and focal seizure onset zone or lack of seizure focus. The de-identified patient data was retrieved from the online International Epilepsy Electrophysiology Portal (IEEG Portal) \cite{ieeg2013}. Detailed information about community detection can be found in the Supplementary Methods.

\subsection*{Effective Connectivity Networks} For each epileptic event, effective connectivity networks were constructed from one-second time windows of both ictal and preictal data. All timeseries data from ECoG channel recordings were re-referenced to the common average reference to avoid broad field effects \cite{Kramer2010, Khambhati2015, Das2019}. Each channel was digitally filtered with 60Hz notch, 120Hz low-pass, and 1Hz high-pass filters using a 4th-order Butterworth design to remove line-noise, drift, and high frequency noise, respectively \cite{Kramer2010}.  The preprocessed data was partitioned into consecutive one-second time windows spanning the length of the seizure and preictal period. For each window, the GLASSO method was used to find a regularized partial correlation EC matrix \cite{Friedman2008}. Detailed information can be found in the Supplementary Methods.

\subsection*{Community Detection to Identify Seizure Phases} A similarity matrix was computed using the Pearson correlation coefficient between all pairwise combinations of the $T$ effective connectivity networks for a single seizure; each element was linearly weighted by a contiguity parameter $\beta= 0.01$, such that networks at two distant time points were weighted less than those at adjacent timepoints. Using MATLAB, the Louvain algorithm from \cite{Blondel2008} was used to maximize a modularity quality function and thereby to identify an assignment of networks to communities from the network similarity matrix. The algorithm aims to iteratively tune an initial set of module boundaries in the similarity network to maximize within-community similarity relative to a null model. The number of communities discovered can be tuned by the scaling parameter $\gamma$. For each EC network, the Louvain algorithm was run 100 times using a value of $\gamma$ that produced 3 communities, and a consensus partition with a corresponding modularity value $Q$ was selected from the 100 trial partitions. Detailed information about community detection can be found in the Supplementary Methods.

\subsection*{Controllability Metrics and Optimal Control Calculations} Average, modal, persistent modal, and transient modal controllability metrics were calculated in MATLAB with custom scripts and code from \cite{Stiso2018}. Before calculating a given controllability metric, each function first ensures that the network is stable by enforcing Schur stability and then subtracting the identity matrix to normalize the eigenvalues \cite{Karrer2019}. Optimal control energy was also calculated in MATLAB using functions referenced in \cite{Stiso2018}. Optimal control energy requires the selection of the value for $\tau$, the number of time steps over which to perform the control trajectory optimization, and $\alpha$, an energy-distance tradeoff parameter. Parameter selection details can be found in the Supplementary Methods. 

\subsection*{Statistical Methods} Friedman’s ANOVA, $t$-tests, and Bonferroni corrections for multiple comparisons were performed using functions from MATLAB’s statistics toolbox. We wrote custom MATLAB scripts to create null distributions for our energy analysis by randomly sampling with replacement 5000 sets of nodes containing the same number of nodes as our set of interest. Our random resampling method is an extension of similar methods used to compute subject-specific confidence intervals for nodal metric values \cite{Conrad2020}.
}

\showmatmethods{} 

\acknow{This work was primarily funded by an NSF award BCS-1631550 to DSB and an NINDS award R01 NS099348 to BL and DSB. DSB would like to acknowledge additional support from the John D. and Catherine T. MacArthur Foundation, the Alfred P. Sloan Foundation, the ISI Foundation, the Paul Allen Foundation, the Army Research Laboratory (W911NF-10-2-0022), and the Army Research Office (Bassett-W911NF-14-1-0679, Grafton-W911NF-16-1-0474). BL would like to acknowledge additional support from the Mirowski Family Foundation, Johnathan \& Bonnie Rothberg, and Neil \& Barbara Smit. The content is solely the responsibility of the authors and does not necessarily represent the official views of any of the funding agencies. 
}

\subsection*{Diversity Statement}
Recent work in neuroscience and other fields has identified a bias in citation practices such that papers from women and other minorities are under-cited relative to the number of such papers in the field \cite{Chakravartty2018,Caplar2017}. We used automatic classification of gender based on the first names of the first and last authors \cite{Dworkin2020}, with possible combinations including male/male, male/female, female/male, and female/female. After excluding self-citations to the first and senior authors of our current paper, the references in this work contain 48.7\% male/male, 23.1\% male/female, 15.4\% female/female, 10.3\% female/male, and 2.6\% unknown citation categorizations. We look forward to future work that could help us better understand how to support equitable practices in science.

\subsection* {Data Accessibility} All data used in this paper was de-identified and can be retrieved from the online International Epilepsy Electrophysiology Portal at \href{https://ieeg.org}{www.ieeg.org}

\subsection*{Author Contributions}
D.S.B. and B.H.S designed research; B.H.S performed research, B.H.S. and A.A. analyzed data, K.A.D and F.M. assisted with data collection, B.H.S, D.S.B. and J.S. wrote the paper, K.A.D, F.P. and B.L. revised the manuscript.

\showacknow{} 

\bibliography{bibliography_v4_2}

\begin{thebibliography}{10}

\bibitem{Bergey2015}
GK Bergey, et~al., {Long-term treatment with responsive brain stimulation in
  adults with refractory partial seizures}.
\newblock {\em\protect\JournalTitle{Neurology}} \textbf{84}, 810--817 (2015).

\bibitem{Hartshorn2018}
A Hartshorn, B Jobst, {Responsive brain stimulation in epilepsy}.
\newblock {\em\protect\JournalTitle{Therapeutic Advances in Chronic Disease}}
  \textbf{9}, 135--142 (2018).

\bibitem{Pasqualetti2014}
F Pasqualetti, S Zampieri, F Bullo, {Controllability metrics, limitations and
  algorithms for complex networks}.
\newblock {\em\protect\JournalTitle{IEEE Transactions on Control of Network
  Systems}} \textbf{1}, 40--52 (2014).

\bibitem{Tang2019}
E Tang, et~al., {The control of brain network dynamics across diverse scales of
  space and time}.
\newblock {\em\protect\JournalTitle{bioRxiv}}, 1--12 (2019).

\bibitem{Muldoon2016}
SF Muldoon, et~al., {Stimulation based control of dynamic brain networks}.
\newblock {\em\protect\JournalTitle{PLoS Computational Biology}} \textbf{12}
  (2016).

\bibitem{Khambhati2018}
AN Khambhati, et~al., {Functional control of electrophysiological network
  architecture using direct neurostimulation in humans}.
\newblock {\em\protect\JournalTitle{Network Neuroscience}}, 1--30 (2019).

\bibitem{Stiso2018}
J Stiso, et~al., {White matter network architecture guides direct electrical
  stimulation through optimal state transitions}.
\newblock {\em\protect\JournalTitle{Cell Reports}} \textbf{28}, 2554--2566.e7
  (2019).

\bibitem{Kim2018}
JZ Kim, et~al., {Role of graph architecture in controlling dynamical networks
  with applications to neural systems}.
\newblock {\em\protect\JournalTitle{Nature Physics}} \textbf{14}, 91--98
  (2018).

\bibitem{Gu2015}
S Gu, et~al., {Controllability of structural brain networks}.
\newblock {\em\protect\JournalTitle{Nature Communications}} \textbf{6}, 1--10
  (2015).

\bibitem{Betzel2016}
RF Betzel, S Gu, JD Medaglia, F Pasqualetti, DS Bassett, {Optimally controlling
  the human connectome: The role of network topology}.
\newblock {\em\protect\JournalTitle{Scientific Reports}} \textbf{6}, 1--14
  (2016).

\bibitem{Smith2011}
SM Smith, et~al., {Network modelling methods for FMRI}.
\newblock {\em\protect\JournalTitle{NeuroImage}} \textbf{54}, 875--891 (2011).

\bibitem{Kramer2012}
MA Kramer, SS Cash, {Epilepsy as a disorder of cortical network organization}.
\newblock {\em\protect\JournalTitle{The Neuroscientist}} \textbf{18}, 360--372
  (2012).

\bibitem{Timofeev2004}
I Timofeev, M Steriade, {Neocortical seizures: Initiation, development and
  cessation}.
\newblock {\em\protect\JournalTitle{Neuroscience}} \textbf{123}, 299--336
  (2004).

\bibitem{Loddenkemper2014}
T Loddenkemper, et~al., {Subunit composition of glutamate and
  gamma-aminobutyric acid receptors in status epilepticus}.
\newblock {\em\protect\JournalTitle{Epilepsy Research}} \textbf{108}, 605--615
  (2014).

\bibitem{Kirst2016}
C Kirst, M Timme, D Battaglia, {Dynamic information routing in complex
  networks}.
\newblock {\em\protect\JournalTitle{Nature Communications}} \textbf{7}, 11061
  (2016).

\bibitem{Palmigiano2017}
A Palmigiano, T Geisel, F Wolf, D Battaglia, {Flexible information routing by
  transient synchrony}.
\newblock {\em\protect\JournalTitle{Nature Neuroscience}} \textbf{20},
  1014--1022 (2017).

\bibitem{Battaglia2012}
D Battaglia, A Witt, F Wolf, T Geisel, {Dynamic effective connectivity of
  inter-areal brain circuits}.
\newblock {\em\protect\JournalTitle{PLoS Computational Biology}} \textbf{8}
  (2012).

\bibitem{Horwitz2003}
B Horwitz, {The elusive concept of brain connectivity}.
\newblock {\em\protect\JournalTitle{NeuroImage}} \textbf{19}, 466--470 (2003).

\bibitem{Sakkalis2011}
V Sakkalis, {Review of advanced techniques for the estimation of brain
  connectivity measured with EEG/MEG}.
\newblock {\em\protect\JournalTitle{Computers in Biology and Medicine}}
  \textbf{41}, 1110--1117 (2011).

\bibitem{Ren2019}
Y Ren, et~al., {Transient seizure onset network for localization of
  epileptogenic zone: effective connectivity and graph theory-based analyses of
  ECoG data in temporal lobe epilepsy}.
\newblock {\em\protect\JournalTitle{Journal of Neurology}} \textbf{266},
  844--859 (2019).

\bibitem{Friedman2008}
J Friedman, T Hastie, R Tibshirani, {Sparse inverse covariance estimation with
  the graphical lasso}.
\newblock {\em\protect\JournalTitle{Biostatistics}} \textbf{9}, 432--441
  (2008).

\bibitem{Rosa2015}
MJ Rosa, et~al., {Sparse network-based models for patient classification using
  fMRI}.
\newblock {\em\protect\JournalTitle{NeuroImage}} \textbf{105}, 493--506 (2015).

\bibitem{Das2019}
A Das, D Sexton, C Lainscsek, SS Cash, TJ Sejnowski, {Characterizing brain
  connectivity from human electrocorticography recordings With unobserved
  inputs during epileptic seizures}.
\newblock {\em\protect\JournalTitle{Neural Computation}} \textbf{31},
  1271--1326 (2019).

\bibitem{Karrer2019}
TM Karrer, et~al., {A practical guide to methodological considerations in the
  controllability of structural brain networks}.
\newblock {\em\protect\JournalTitle{Journal of Neural Engineering}}, 1--33
  (2020).

\bibitem{shine2019human}
JM Shine, et~al., Human cognition involves the dynamic integration of neural
  activity and neuromodulatory systems.
\newblock {\em\protect\JournalTitle{Nature Neuroscience}} \textbf{22}, 289--296
  (2019).

\bibitem{Wu-Yan2018}
E Wu-Yan, et~al., {Benchmarking measures of network controllability on
  canonical graph models}.
\newblock {\em\protect\JournalTitle{Journal of Nonlinear Science}}, 1--39
  (2018).

\bibitem{Khambhati2016}
AN Khambhati, KA Davis, TH Lucas, B Litt, DS Bassett, {Virtual cortical
  resection reveals push-pull network control preceding seizure Eeolution}.
\newblock {\em\protect\JournalTitle{Neuron}} (2016).

\bibitem{Khambhati2015}
AN Khambhati, et~al., {Dynamic network drivers of seizure generation,
  propagation and termination in human neocortical epilepsy}.
\newblock {\em\protect\JournalTitle{PLoS Computational Biology}} (2015).

\bibitem{Jiruska2013}
P Jiruska, et~al., {Synchronization and desynchronization in epilepsy:
  controversies and hypotheses}.
\newblock {\em\protect\JournalTitle{The Journal of Physiology}} \textbf{591},
  787--797 (2013).

\bibitem{Morrell2011}
MJ Morrell, {Responsive cortical stimulation for the treatment of medically
  intractable partial epilepsy}.
\newblock {\em\protect\JournalTitle{Neurology}} \textbf{77}, 1295--1304 (2011).

\bibitem{Weiss2015}
SA Weiss, et~al., {Seizure localization using ictal phase-locked high gamma: A
  retrospective surgical outcome study}.
\newblock {\em\protect\JournalTitle{Neurology}} \textbf{84}, 2320--2328 (2015).

\bibitem{Weiss2013}
SA Weiss, et~al., {Ictal high frequency oscillations distinguish two types of
  seizure territories in humans}.
\newblock {\em\protect\JournalTitle{Brain}} \textbf{136}, 3796--3808 (2013).

\bibitem{Baldassano2017}
SN Baldassano, et~al., {Crowdsourcing seizure detection: algorithm development
  and validation on human implanted device recordings}.
\newblock {\em\protect\JournalTitle{Brain}} \textbf{140}, 1680--1691 (2017).

\bibitem{Kramer2010}
Ma Kramer, et~al., {Coalescence and fragmentation of cortical networks during
  focal seizures.}
\newblock {\em\protect\JournalTitle{The Journal of neuroscience : the official
  journal of the Society for Neuroscience}} \textbf{30}, 10076--85 (2010).

\bibitem{Proix2018}
T Proix, VK Jirsa, F Bartolomei, M Guye, W Truccolo, {Predicting the
  spatiotemporal diversity of seizure propagation and termination in human
  focal epilepsy}.
\newblock {\em\protect\JournalTitle{Nature Communications}} \textbf{9} (2018).

\bibitem{Shah2019}
P Shah, et~al., {Characterizing the role of the structural connectome in
  seizure dynamics}.
\newblock {\em\protect\JournalTitle{Brain}} \textbf{142}, 1955--1972 (2019).

\bibitem{Kandel2000}
E Kandel, J Schwartz, T Jessell, {The modulatory functions of the brain stem}
  in {\em Principles of Neural Science}.
\newblock (McGraw-Hill, Health Professions Division, New York), 6th edition,
  pp. 1038--1055 (2000).

\bibitem{Mullen2012}
T Mullen, G Worrell, S Makeig, {Multivariate principal oscillation pattern
  analysis of ICA sources during seizure} in {\em 2012 Annual International
  Conference of the IEEE Engineering in Medicine and Biology Society}.
\newblock (IEEE), pp. 2921--2924 (2012).

\bibitem{Lehnertz2016}
K Lehnertz, H Dickten, S Porz, C Helmstaedter, CE Elger, {Predictability of
  uncontrollable multifocal seizures - Towards new treatment options}.
\newblock {\em\protect\JournalTitle{Scientific Reports}} \textbf{6}, 1--9
  (2016).

\bibitem{Spencer2002}
SS Spencer, {Neural networks in human epilepsy: evidence of and implications
  for treatment}.
\newblock {\em\protect\JournalTitle{Epilepsia}} \textbf{43}, 219--227 (2002).

\bibitem{Ung2017}
H Ung, et~al., {Interictal epileptiform activity outside the seizure onset zone
  impacts cognition}.
\newblock {\em\protect\JournalTitle{Brain}} \textbf{140}, 2157--2168 (2017).

\bibitem{Seeck2013}
M Seeck, et~al., {Electrode location and clinical outcome in hippocampal
  electrical stimulation for mesial temporal lobe epilepsy}.
\newblock {\em\protect\JournalTitle{Seizure}} \textbf{22}, 390--395 (2013).

\bibitem{Morari1988}
M Morari, CE Garcia, DM Prett, {Model predictive control: Theory and practice}.
\newblock {\em\protect\JournalTitle{IFAC Proceedings Volumes}} \textbf{21},
  1--12 (1988).

\bibitem{Ashourvan2020}
A Ashourvan, et~al., {Model-based design for seizure control by stimulation}.
\newblock {\em\protect\JournalTitle{Journal of Neural Engineering}}, 0--11
  (2020).

\bibitem{Wild2010}
B Wild, et~al., {A graphical vector autoregressive modelling approach to the
  analysis of electronic diary data}.
\newblock {\em\protect\JournalTitle{BMC Medical Research Methodology}}
  \textbf{10}, 28 (2010).

\bibitem{Epskamp2018}
S Epskamp, EI Fried, {A tutorial on regularized partial correlation networks}.
\newblock {\em\protect\JournalTitle{Psychological Methods}} \textbf{23},
  617--634 (2018).

\bibitem{Colombo2019}
MA Colombo, et~al., {The spectral exponent of the resting EEG indexes the
  presence of consciousness during unresponsiveness induced by propofol, xenon,
  and ketamine}.
\newblock {\em\protect\JournalTitle{NeuroImage}} \textbf{189}, 631--644 (2019).

\bibitem{Schaworonkow2019}
N Schaworonkow, VV Nikulin, {Spatial neuronal synchronization and the waveform
  of oscillations: Implications for EEG and MEG}.
\newblock {\em\protect\JournalTitle{PLOS Computational Biology}} \textbf{15},
  e1007055 (2019).

\bibitem{Zanudo2017}
JGT Za{\~{n}}udo, G Yang, R Albert, {Structure-based control of complex
  networks with nonlinear dynamics}.
\newblock {\em\protect\JournalTitle{Proceedings of the National Academy of
  Sciences}} \textbf{114}, 7234--7239 (2017).

\bibitem{ieeg2013}
JB Wagenaar, BH Brinkmann, Z Ives, GA Worrell, B Litt, A multimodal platform
  for cloud-based collaborative research in {\em 2013 6th International
  IEEE/EMBS Conference on Neural Engineering (NER)}.
\newblock pp. 1386--1389 (2013).

\bibitem{Blondel2008}
VD Blondel, JL Guillaume, R Lambiotte, E Lefebvre, {Fast unfolding of
  communities in large networks}.
\newblock {\em\protect\JournalTitle{Journal of Statistical Mechanics: Theory
  and Experiment}} \textbf{2008}, P10008 (2008).

\bibitem{Conrad2020}
EC Conrad, et~al., {The sensitivity of network statistics to incomplete
  electrode sampling on intracranial EEG}.
\newblock {\em\protect\JournalTitle{Network Neuroscience}}, 1--52 (2020).

\bibitem{Chakravartty2018}
P Chakravartty, R Kuo, V Grubbs, C McIlwain, {{\#}CommunicationSoWhite}.
\newblock {\em\protect\JournalTitle{Journal of Communication}} \textbf{68},
  254--266 (2018).

\bibitem{Caplar2017}
N Caplar, S Tacchella, S Birrer, {Quantitative evaluation of gender bias in
  astronomical publications from citation counts}.
\newblock {\em\protect\JournalTitle{Nature Astronomy}} \textbf{1} (2017).

\bibitem{Dworkin2020}
JD Dworkin, et~al., {The extent and drivers of gender imbalance in neuroscience
  reference lists}.
\newblock {\em\protect\JournalTitle{arXiv}} (2020).

\end{thebibliography}


\begin{thebibliography}{10}

\bibitem{Gu2015}
S Gu, et~al., {Controllability of structural brain networks}.
\newblock {\em\protect\JournalTitle{Nature Communications}} \textbf{6}, 8414
  (2015).

\bibitem{Pasqualetti2014}
F Pasqualetti, S Zampieri, F Bullo, {Controllability metrics, limitations and
  algorithms for complex networks}.
\newblock {\em\protect\JournalTitle{IEEE Transactions on Control of Network
  Systems}} \textbf{1}, 40--52 (2014).

\bibitem{Muldoon2016}
SF Muldoon, et~al., {Stimulation-based control of dynamic brain networks}.
\newblock {\em\protect\JournalTitle{PLoS Computational Biology}} \textbf{12},
  e1005076 (2016).

\bibitem{Galan2008}
RF Gal{\'{a}}n, {On how network architecture determines the dominant patterns
  of spontaneous neural activity}.
\newblock {\em\protect\JournalTitle{PLoS ONE}} \textbf{3}, e2148 (2008).

\bibitem{Stiso2018}
J Stiso, et~al., {White matter network architecture guides direct electrical
  stimulation through optimal state transitions}.
\newblock {\em\protect\JournalTitle{Cell Reports}} \textbf{28}, 2554--2566.e7
  (2019).

\bibitem{Khambhati2018}
AN Khambhati, et~al., {Functional control of electrophysiological network
  architecture using direct neurostimulation in humans}.
\newblock {\em\protect\JournalTitle{Network Neuroscience}} \textbf{3}, 848--877
  (2018).

\bibitem{Kirst2016}
C Kirst, M Timme, D Battaglia, {Dynamic information routing in complex
  networks}.
\newblock {\em\protect\JournalTitle{Nature Communications}} \textbf{7}, 11061
  (2016).

\bibitem{Palmigiano2017}
A Palmigiano, T Geisel, F Wolf, D Battaglia, {Flexible information routing by
  transient synchrony}.
\newblock {\em\protect\JournalTitle{Nature Neuroscience}} \textbf{20},
  1014--1022 (2017).

\bibitem{Wu-Yan2018}
E Wu-Yan, et~al., {Benchmarking measures of network controllability on
  canonical graph models}.
\newblock {\em\protect\JournalTitle{Journal of Nonlinear Science}}, 1--39
  (2018).

\bibitem{Jiruska2013}
P Jiruska, et~al., {Synchronization and desynchronization in epilepsy:
  controversies and hypotheses}.
\newblock {\em\protect\JournalTitle{The Journal of Physiology}} \textbf{591},
  787--797 (2013).

\bibitem{Kramer2012}
MA Kramer, SS Cash, {Epilepsy as a disorder of cortical network organization}.
\newblock {\em\protect\JournalTitle{Neuroscientist}} \textbf{18}, 360--372
  (2012).

\bibitem{Kramer2010}
MA Kramer, et~al., {Coalescence and fragmentation of cortical networks during
  focal seizures}.
\newblock {\em\protect\JournalTitle{Journal of Neuroscience}} \textbf{30},
  10076--10085 (2010).

\bibitem{Proix2018}
T Proix, VK Jirsa, F Bartolomei, M Guye, W Truccolo, {Predicting the
  spatiotemporal diversity of seizure propagation and termination in human
  focal epilepsy}.
\newblock {\em\protect\JournalTitle{Nature Communications}} \textbf{9} (2018).

\bibitem{Chen2008}
J Chen, Z Chen, {Extended Bayesian information criteria for model selection
  with large model spaces}.
\newblock {\em\protect\JournalTitle{Biometrika}} \textbf{95}, 759--771 (2008).

\bibitem{Foygel2010}
R Foygel, M Drton, {Extended Bayesian information criteria for Gaussian
  graphical models}.
\newblock {\em\protect\JournalTitle{Advances in Neural Information Processing
  Systems 23: 24th Annual Conference on Neural Information Processing Systems
  2010, NIPS 2010}}, 1--14 (2010).

\bibitem{Karrer2019}
TM Karrer, et~al., {A practical guide to methodological considerations in the
  controllability of structural brain networks}.
\newblock {\em\protect\JournalTitle{Journal of Neural Engineering}}, 1--33
  (2020).

\bibitem{Epskamp2018}
S Epskamp, EI Fried, {A tutorial on regularized partial correlation networks}.
\newblock {\em\protect\JournalTitle{Psychological Methods}} \textbf{23},
  617--634 (2018).

\bibitem{Banerjee2007}
O Banerjee, LE Ghaoui, A D'Aspremont, {Model selection through sparse maximum
  likelihood estimation}.
\newblock {\em\protect\JournalTitle{J. Machine Learning Research}} (2007).

\bibitem{Friedman2008}
J Friedman, T Hastie, R Tibshirani, {Sparse inverse covariance estimation with
  the graphical lasso}.
\newblock {\em\protect\JournalTitle{Biostatistics}} \textbf{9}, 432--441
  (2008).

\bibitem{colclough2018}
GL Colclough, et~al., {Multi-subject hierarchical inverse covariance modelling
  improves estimation of functional brain networks}.
\newblock {\em\protect\JournalTitle{NeuroImage}} \textbf{178}, 370--384 (2018).

\bibitem{Rosa2015}
MJ Rosa, et~al., {Sparse network-based models for patient classification using
  fMRI}.
\newblock {\em\protect\JournalTitle{NeuroImage}} \textbf{105}, 493--506 (2015).

\bibitem{Smith2011}
SM Smith, et~al., {Network modelling methods for FMRI}.
\newblock {\em\protect\JournalTitle{NeuroImage}} \textbf{54}, 875--891 (2011).

\bibitem{Wang2018}
Q Wang, et~al., {EECoG-Comp: An open source platform for concurrent EEG/ECoG
  comparisons: applications to connectivity studies}.
\newblock {\em\protect\JournalTitle{bioRxiv}}, 350199 (2018).

\bibitem{ZhaoHaoming2019}
XL {Zhao, Haoming Jiang, Xinyu Fei, Han Liu, Kathryn Roeder, John Lafferty,
  Larry Wasserman}, {huge: high-dimensional undirected graph estimation}
  (2019).

\bibitem{Liu2009}
H Liu, J Lafferty, L Wasserman, {The nonparanormal: Semiparametric estimation
  of high dimensional undirected graphs}.
\newblock {\em\protect\JournalTitle{Journal of Machine Learning Research}}
  \textbf{10}, 2295--2328 (2009).

\bibitem{Blondel2008}
VD Blondel, JL Guillaume, R Lambiotte, E Lefebvre, {Fast unfolding of
  communities in large networks}.
\newblock {\em\protect\JournalTitle{Journal of Statistical Mechanics: Theory
  and Experiment}} \textbf{2008}, P10008 (2008).

\bibitem{Girvan2002}
M Girvan, MEJ Newman, {Community structure in social and biological networks}.
\newblock {\em\protect\JournalTitle{Proceedings of the National Academy of
  Sciences}} \textbf{99}, 7821--7826 (2002).

\bibitem{Reichardt2006}
J Reichardt, S Bornholdt, {Statistical mechanics of community detection}.
\newblock {\em\protect\JournalTitle{Physical Review E - Statistical, Nonlinear,
  and Soft Matter Physics}} \textbf{74}, 1--14 (2006).

\bibitem{Tang2019}
E Tang, et~al., {The control of brain network dynamics across diverse scales of
  space and time}.
\newblock {\em\protect\JournalTitle{arXiv}}, 1--12 (2019).

\bibitem{Weiss2013}
SA Weiss, et~al., {Ictal high frequency oscillations distinguish two types of
  seizure territories in humans}.
\newblock {\em\protect\JournalTitle{Brain}} \textbf{136}, 3796--3808 (2013).

\bibitem{Baldassano2017}
SN Baldassano, et~al., {Crowdsourcing seizure detection: algorithm development
  and validation on human implanted device recordings}.
\newblock {\em\protect\JournalTitle{Brain}} \textbf{140}, 1680--1691 (2017).

\bibitem{Kohling2000}
R K{\"{o}}hling, M Vreugdenhil, E Bracci, JGR Jefferys, {Ictal epileptiform
  activity is facilitated by hippocampal GABA A receptor-mediated
  oscillations}.
\newblock {\em\protect\JournalTitle{The Journal of Neuroscience}} \textbf{20},
  6820--6829 (2000).

\bibitem{Weiss2015}
SA Weiss, et~al., {Seizure localization using ictal phase-locked high gamma: A
  retrospective surgical outcome study}.
\newblock {\em\protect\JournalTitle{Neurology}} \textbf{84}, 2320--2328 (2015).

\end{thebibliography}

\end{document}


\maketitle

\SItext
\subsection*{Interpretation of dynamic connectivity and controllability}
Prior work \cite{Gu2015,Pasqualetti2014, Muldoon2016, Galan2008} has demonstrated the utility of approximating the evolution of brain dynamics in a network as a linear system governed by 
\begin{equation}
\dot{x}(t)=\boldsymbol{A}x(t)+\boldsymbol{B}_K u_K (t), \label{eq1}
\end{equation}
where $x\in R^N$ is a vector representing brain state, $\boldsymbol{A}\in R^{NxN}$ is the adjacency matrix describing network connectivity, $u_K\in R^m$ is a function of control input over time, and $\boldsymbol{B}_K\in R^{Nxm}$ is the control input matrix identifying the set of control points $K$ where
\begin{equation}
	K={\{k_1,\ldots,k_m }\},\ \boldsymbol{B}_K=[e_{k_i }\ldots e_{k_m } ], \label{eq2}
\end{equation}
in which $e_i$ denotes the $i^{th}$ canonical vector of dimension $N$. In studies of brain control using this linear framework, $\boldsymbol{A}$ is typically assumed to be static over time, and is based on a structural connectivity network with purely positive elements \cite{Gu2015, Stiso2018, Khambhati2018}.  In this work we ease the static assumption by introducing multi-phase, signed connectivity, such that
\begin{equation}
	\dot{x}(t)=\boldsymbol{A}_p x(t)+\boldsymbol{\boldsymbol{B}}_k u_k (t),	\label{eq3}
\end{equation}
where $\boldsymbol{A}_p$ is the representative effective connectivity matrix for a given seizure phase $p$. Here we take the effective connectivity networks, $\boldsymbol{EC_{net}}$, built from one-second time windows throughout the seizure and ensure stability of each network by using the weighted Laplacian as our connectivity matrix $\boldsymbol{A}$: 
\begin{equation}
	\boldsymbol{A}=\frac{\boldsymbol{EC_{net}}}{1+\lambda_{max}}-\boldsymbol{I},
\end{equation}
where $\lambda_{max}$ is the largest eigenvalue of the given EC network, and $\boldsymbol{I}$ is the identity matrix. Next, we assign each stabilized network to a seizure phase. The set of seizure phases is determined by using community detection to group together the networks that share a similar pattern of edge weights (see subsection on ``Modularity maximization of similarity matrices'' below). When the community detection algorithm is tuned to find three network communities, the grouped networks generally fall into consecutive time periods, or seizure phases, corresponding to seizure onset, propagation, and termination.

It has been theorized that information propagation through the brain may depend on the underlying state of synchrony between brain regions, and furthermore that exogenous inputs to brain regions may bias the direction of information flow \cite{Kirst2016, Palmigiano2017}. The switched linear system model above represents a first approximation of continuously transitioning effective connectivity, where each seizure phase defines a distinct pattern of corticocortial coupling. Linear control theory can be applied within each seizure phase to provide time-varying controllability metrics. Thus, controllability here refers to the ability to drive neural interactions while the brain is exhibiting a particular pattern of cortical influences.

\section*{Supplementary Results}
\subsection*{Strength of EC networks across phases}
\begin{figure}
\centering
\includegraphics[width=0.65\textwidth]{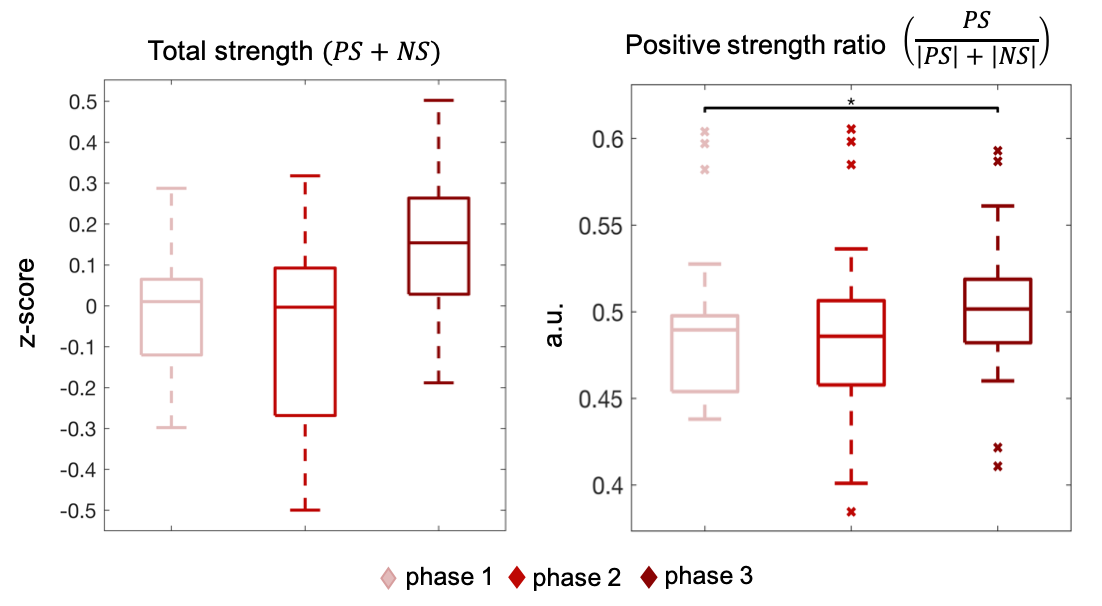}
\caption{\textbf{Group-level network strength across ictal phases.} A significant effect of seizure phase on network strength was found ($\chi^2(2, 34)=7.82, p<0.02$), although \emph{post-hoc} testing did not uncover any significant differences between phases. Total network strength was calculated as the sum of all positive and negative edges in a network. With respect to the sign of the edge weights, a significant increase in positive strength normalized by the total absolute value of all edges weights was found from onset to termination phases. Boxplots indicate the 75\% confidence interval (box), median (solid line), 95\% confidence interval (whiskers), and outliers (stars). Starred bars indicate significant metric differences between phases using the $t$-test at the $p<0.017$ level. PS- total positive edge weights, NS- total negative edge weights.}
\label{fig:strength}
\end{figure}

We begin by noting that analysis of controllability metrics on \emph{structural} brain networks has revealed that the network topology gives rise to a positive (negative) correlation between average (modal) controllability and the rank of the weighted degree, or node strength \cite{Gu2015, Wu-Yan2018, Muldoon2016}. We therefore hypothesized that dynamic reorganization and increased \emph{effective connectivity} strength during seizure propagation would accompany the increase in average controllability and decrease in modal controllability seen after seizure onset \cite{Gu2015}. To test our hypothesis, in all EC networks we calculated node strength as the sum of all positive and negative edges connecting to a node. We then found the median metric value within a seizure phase to obtain a representative phase-specific measure of node strength. At the group level, a significant effect of seizure phase on network strength was found ($\chi^2(2, 34)=7.82, p<0.02$). Strength increased from onset to termination, although \emph{post-hoc} testing did not uncover any significant differences in node strength between consecutive phase pairs. We also inspected the change in the proportion of positive edge weight values during seizures. For a given network we first calculated the magnitude of positive edge weights normalized by the sum of the absolute values of all edge weights for a node, and then averaged across nodes. We chose the median network strength ratio from all networks assigned to a phase, to represent that seizure phase. During seizure onset for the median seizure, less than 50\% of absolute nodal strength was comprised of positive edge weights, and the positive edge strength ratio rose significantly in the termination phase compared with seizure onset ($N=34, t=-3.03 , p<0.008$) \ref{fig:strength}. The increase in network strength from seizure onset to termination found throughout the majority of the seizures in our study is consistent with the observed phenomenon of increasing synchrony over the ictal period \cite{Jiruska2013,Kramer2012}.
 
\subsection*{Controllability across phases in preictal data}
\begin{figure}
\centering
\includegraphics[width=0.8\textwidth]{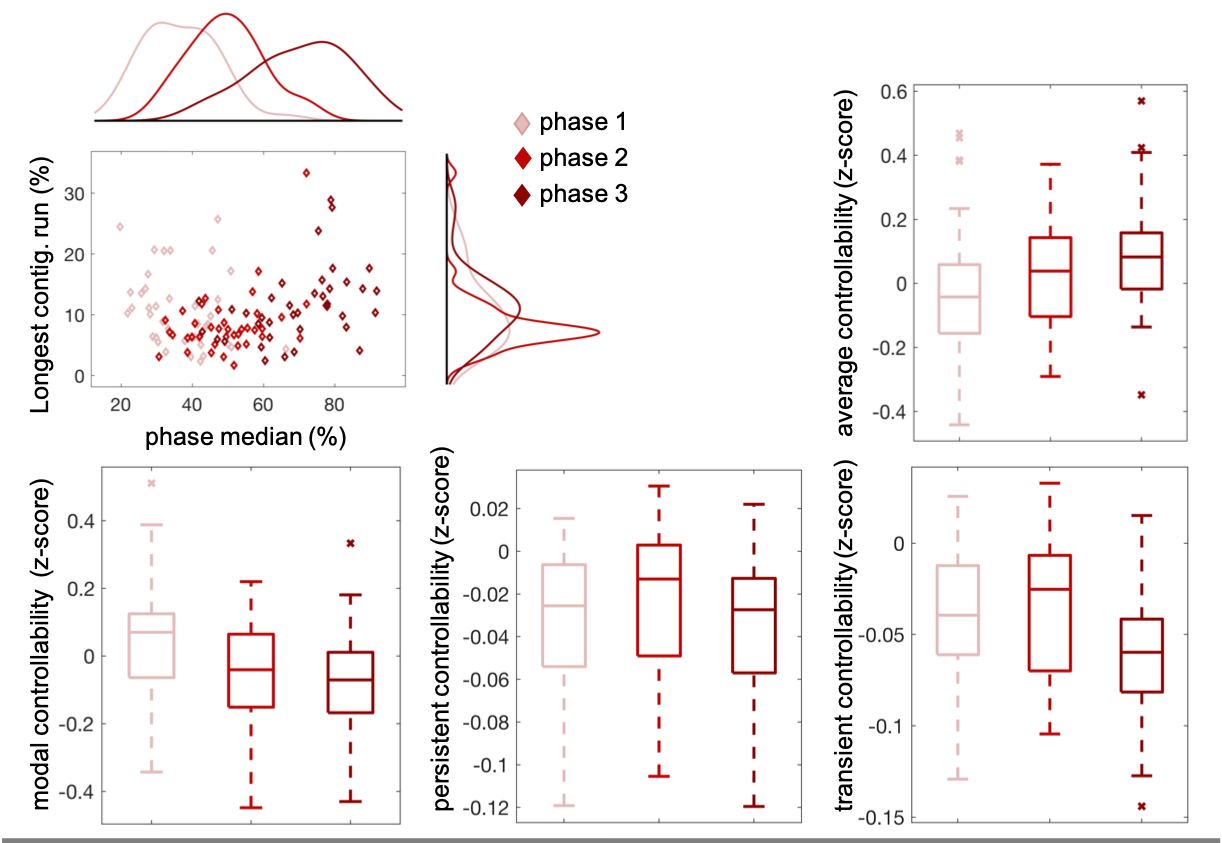}
\caption{\textbf{Preictal network community detection and metric calculations.} We maximized the modularity of a weighted matrix encoding the similarity among all preictal effective connectivity networks ($n=34$) before a seizure. This process produces an assignment of each preictal time window to one of three communities. We labeled each community chronologically as phases one through three by finding the median of the longest temporally contiguous community assignment (\emph{upper left}). Unlike the ictal seizure phases, no significant effect of community classification was found on any of the controllability metrics, assessed using Friedman’s ANOVA and using a Bonferroni corrected \emph{post-hoc} $t$-test significance level of $p<0.017$ to correct for multiple comparisons across phases in a seizure. Boxplots indicate the 75\% confidence interval (box), median (solid line), 95\% confidence interval (whiskers), and outliers (stars). }
\label{fig:preictal}
\end{figure}

In our main work, we compared network controllability values across three seizure phases determined using a data-driven modularity maximization method. To ensure that significant differences between network metrics in each phase were not simply driven by underlying variance of network metrics observed at baseline, here we perform the same community detection and metric comparison on EC networks built from preictal data comprised of time series equal in length to the subsequent seizure (Fig \ref{fig:preictal}). The preictal period does not commonly display distinct phases, and thus we do not expect a significant difference in metric values between similarity-driven groups of interictal networks. Consistent with our expectations, a group level analysis comparing metrics between three interictal similarity communities before each seizure shows no significant difference in average ($\chi^2(2,34)=3.35,p=0.19)$, modal ($\chi^2 (2,34)=4.65,p=0.1)$, persistent modal ($\chi^2(2,34)=5.35, p=0.1$) or transient modal ($\chi^2 (2,34)=1.24,p=0.54$) controllability metrics (Fig. \ref{fig:preictal}). These results on preictal data indicate that significant differences between phase metrics are specific to the ictal period and cannot be fully explained by an underlying variation in network structure present outside of the ictal period. 

\subsection*{Group-Level Seizure Independence}
\begin{figure}
\centering
\includegraphics[width=\textwidth]{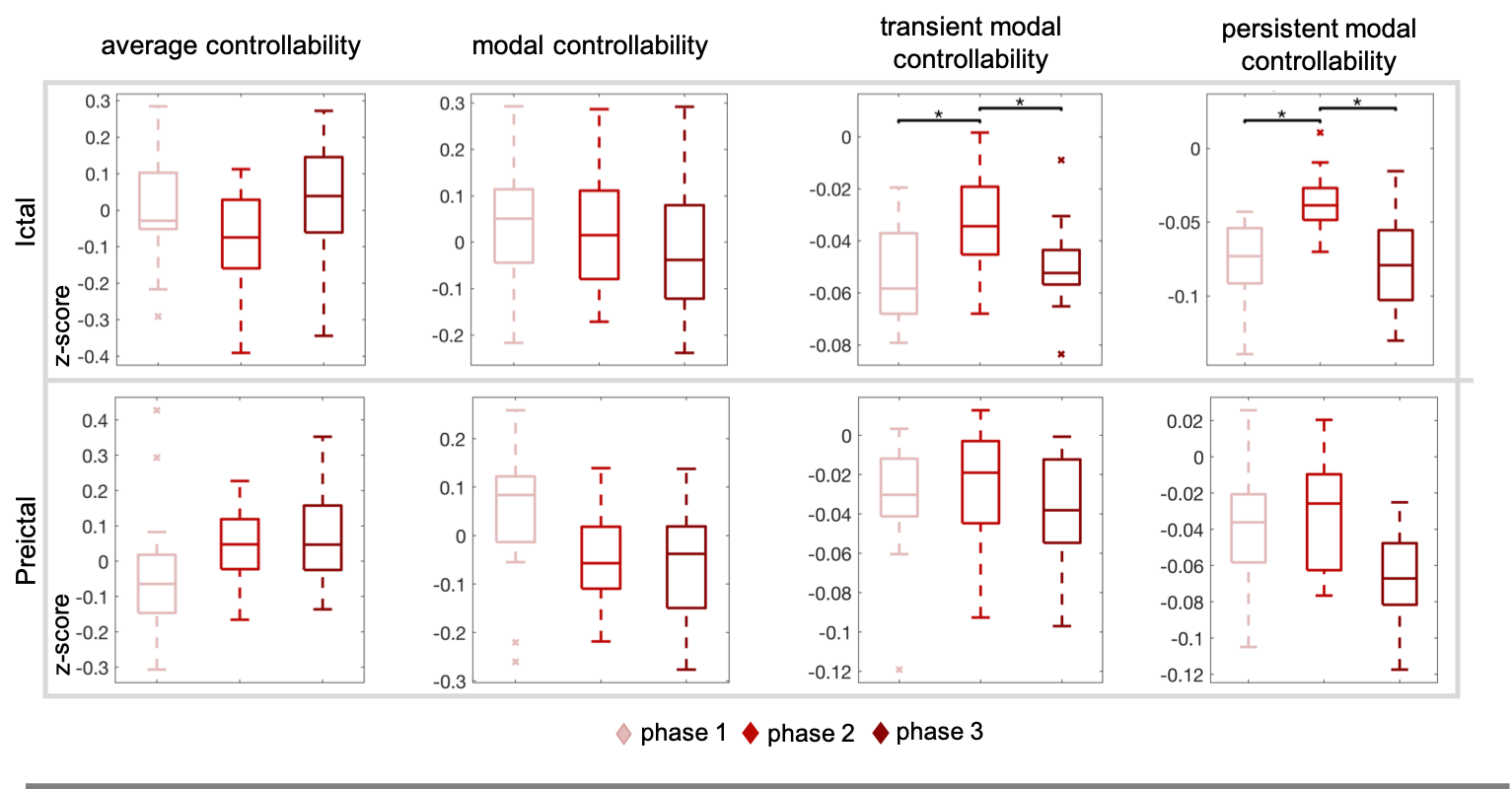}
\caption{\textbf{Group-level phase metrics after stereotyped grouping.} In an alternative analysis to our main group-level results, we did not treat each seizure as an independent data point, and instead averaged metrics between seizures that originated from the same onset zone, treating them as a single data point. Boxplots indicate the 75\% confidence interval (box), median (solid line), 95\% confidence interval (whiskers), and outliers (stars). Starred bars indicate significant metric differences between phases at the $p<0.017$ level, determined using Friedman's ANOVA.}
\label{fig:grouped}
\end{figure}

In our group-level analysis, we treated each seizure as an independent data point to preserve any variance introduced by the diversity of spatiotemporal patterns that may arise between seizures in a single subject \cite{Kramer2010, Proix2018}. However, because seizures in some patients may be highly stereotyped and could bias the group-level results, here we perform the analysis again after grouping seizures originating from the same onset zone within each patient, and averaging phase metrics for those seizures. All seizures in 7 individuals originated from a single onset zone; three patients displayed two onset zones, and the final four additional patients displayed diffuse onsets. Collectively, we therefore considered a total of 17 independent data points. Figure \ref{fig:grouped} shows ictal and preictal results after averaging phase metrics across stereotyped seizures. Consistent with the results described in the section above, there was no significant difference between preictal phases for any metric after grouping: average ($\chi^2(2, 17)=3.38, p=0.19$), modal ($\chi^2(2, 17)=3.38, p=0.19$), transient modal ($\chi^2(2, 17)=2.38, p=0.31$), and persistent modal ($\chi^2(2, 17)=2, p=0.34$) controllability. The ictal trends were similar to our main results after the onset grouping, with phase differences in transient modal ($\chi^2(2, 17)=14, p=0.001$), and persistent modal ($\chi^2(2, 17)=13.5, p=0.001$) controllability maintaining statistical  significance. Average ($\chi^2(2, 17)=3.38, p=0.19$) and modal ($\chi^2(2, 17)=0.38, p=0.83$) controllability were not statistically significant, although the ranking of median phase values did not change. Our alternative analysis demonstrates that the group-level controllability trends remain similar, whether seizures are treated as independent events or as stereotyped onset events within a patient. 

\subsection*{Subject-level analyses}
\begin{figure}
\centering
\includegraphics[width=0.65\textwidth]{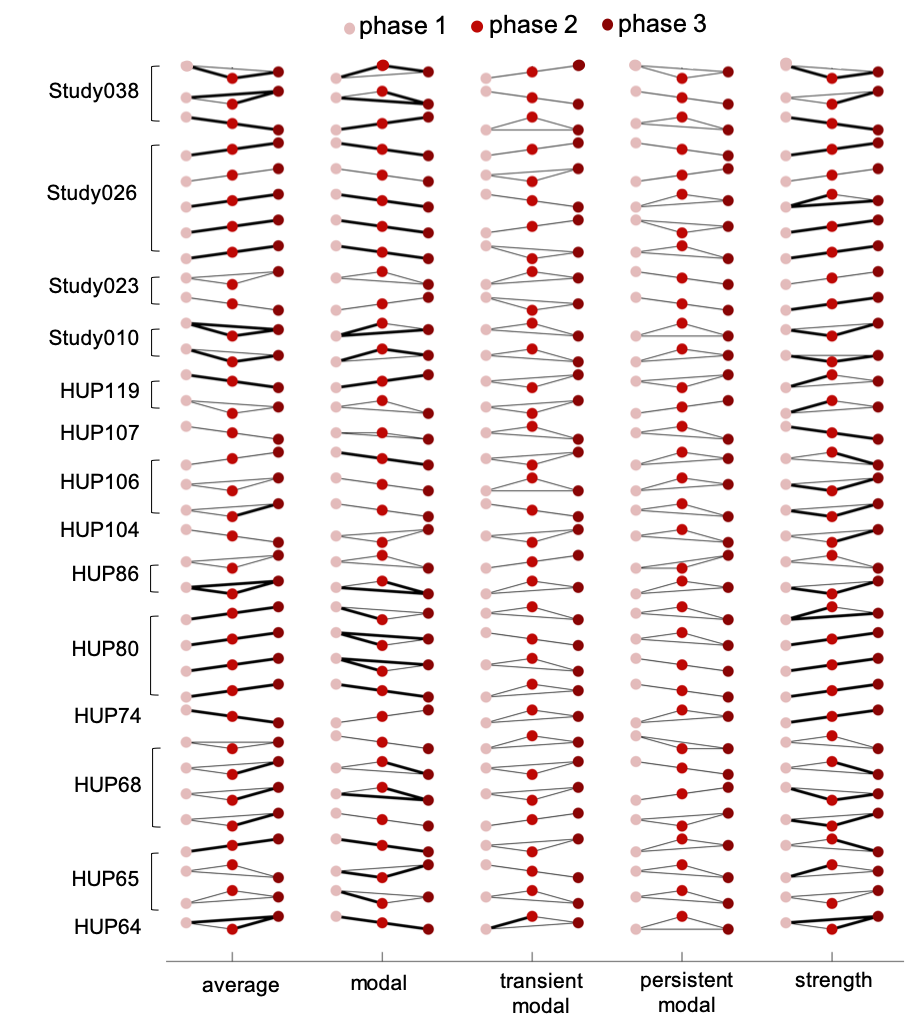}
\caption{\textbf{Dependence of metrics on seizure phase.} Each closed group of three data points represents relative phase metric means from a single seizure. The connecting lines between phases highlight the relative increase or decrease in the mean of the given controllability or strength metric calculated across nodes in each phase network. Bold lines indicate that differences in nodal controllability means between the connected phases were significant. Significance was determined using Friedman’s ANOVA, with a Bonferroni corrected \emph{post-hoc} $t$-test significance level of $p<0.017$ to account for multiple comparisons of phases within the same seizure.}
\label{fig:subject_level}
\end{figure}

We primarily discussed group-level trends of controllability metrics between seizure phases in our main manuscript to obtain a generalizable result. Here we investigate how controllability metrics at the individual seizure level may vary from, or contribute to, the group-level result. For every seizure, we calculated controllability metrics across each node in each EC network, and then averaged nodal values within each seizure phase. We tested whether seizure phase had a significant impact on nodal mean metrics using Friedman’s ANOVA, with a Bonferroni corrected \emph{post-hoc} $t$-test significance level of $p<0.017$ to account for multiple comparisons of phases within the same seizure (Figure \ref{fig:subject_level}). We found that none of the phase differences in transient or persistent modal controllability values were significant in a single seizure, and the heterogeneity in phase means across seizures of an individual reinforced the choice to treat each seizure as an independent event. We found that 73\% of the seizures displayed an increase in average controllability between propagation and termination phases, 55\% significantly so, and 61\% of seizures displayed a decrease in modal controllability between onset and termination phases, 35\% significantly. This finding indicates that the group level results that we reported are representative for most seizures, and that they are not being driven by a small but strong few. 

\subsection*{Robustness of main results to parameter selection}
The methods we used for our analysis required us to choose values for multiple parameters. We selected two parameters throughout the EC network generation and community detection pipeline, and three additional parameters in our controllability metrics and optimal control energy analysis. Here we assess the sensitivity of our main findings to a range of values for the study parameters to ensure robustness of our results. We use a subset of 14 seizures, drawn randomly from each patient, respectively, as our robustness dataset.\\

\begin{figure}
\centering
\includegraphics[width=0.8\textwidth]{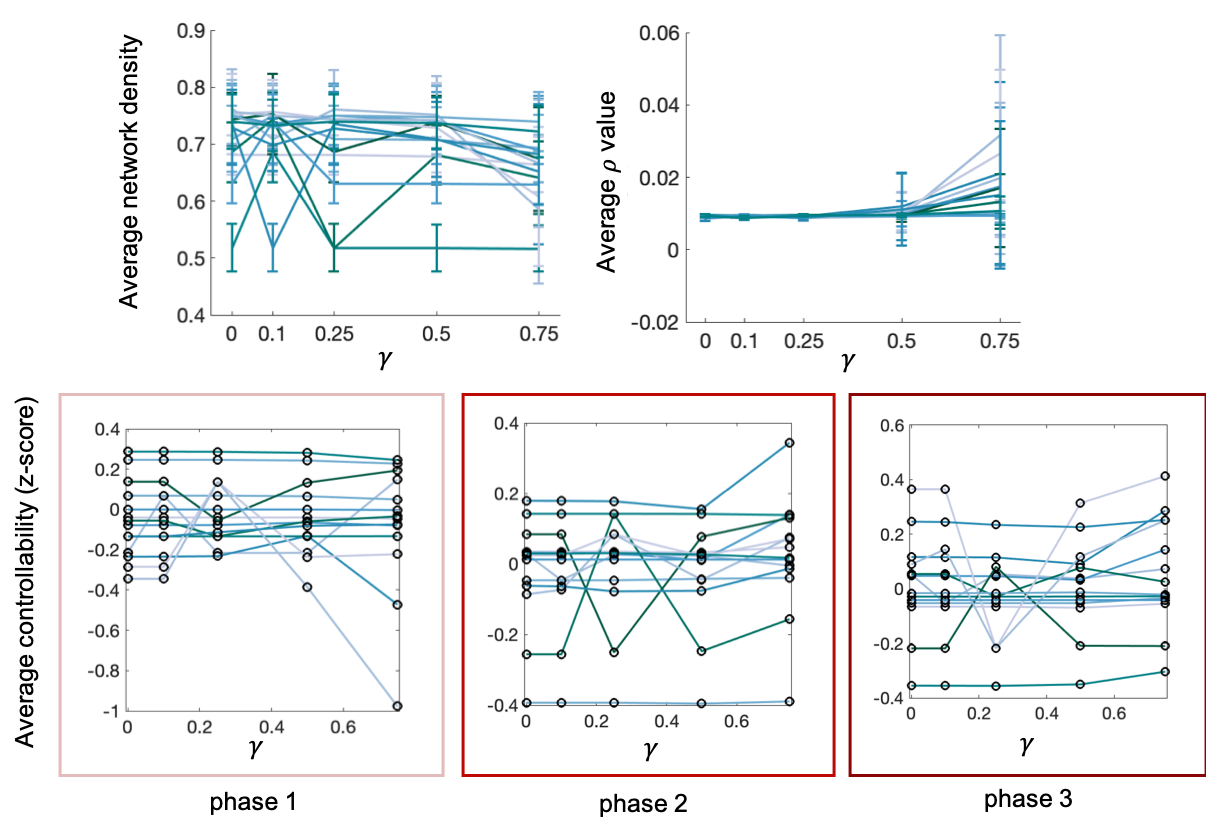}
\caption{\textbf{Variation of EC network sparsity.} The parameter $\lambda$ controls the false positive tolerance when estimating a regularized partial correlation matrix. \emph{Top row}: The average EC network density throughout a seizure and average value for the regularization parameter $\rho$ are shown. Each line represents a single seizure in the robustness dataset $(n=14)$; error bars indicate standard deviation and different line colors are used for ease of visualization \emph{Bottom row}: Average controllability values for each seizure with increasing EC network sparsity. Each line represents a single seizure in the robustness dataset, with each marker representing the network average controllability value calculated from networks generated at the given ${\gamma}$ value. Each panel represents the onset, propagation, and termination phases, respectively.
}
\label{fig:gamma}
\end{figure}

\noindent
\emph{Extended Bayesian Information Criterion (EBIC) hyperparameter $\lambda$:} We used the GLASSO method to estimate effective connectivity matrices from one-second windows of ECoG data (see the later subsection on ``Estimation of effective connectivity'' in the Methods section of this Supplement). In this method, the sparsity of the estimated inverse covariance matrix increases with larger values of the regularization parameter $\rho$. One method of determining the appropriate sparsity level is to evaluate the significance of each edge, and to remove edges that do not meet an appropriate significance criteria. However, when many edges are present, the significance criteria may become stringent after correcting for multiple pairwise comparisons, resulting in the removal of a number of true edges. Alternatively, the network that satisfies the extended Bayesian Information Criterion (EBIC) for model selection will satisfy a desired balance between the exploration of new edges, and the false discovery of spurious edges \cite{Chen2008, Foygel2010}. EBIC relies on a single hyperparameter, $\lambda$, to select either simpler models with fewer edges or more exploratory models with a greater chance of false discovery. In our main work we choose a conservative value of $\lambda=0.5$, at the end of the standard parameter range that prefers a low false discovery rate at the cost of missing some true edges \cite{Foygel2010}. We demonstrate the effect of $\lambda$ (across the value set {0, 0.1, 0.5, 0.75}) on network density, community detection, and controllability values in the robustness dataset ($n=14$) in Figure \ref{fig:gamma}. For each controllability metric, we used the Wilcoxon signed rank test to make pairwise comparisons between the metric values of phase networks for pairs of $\lambda$ values. We found no significant difference in controllability values for any metric or phase between $\lambda$=0.5 and $\lambda$= 0.25 ($n=14, W \leq 54, p > 0.19$), indicating that our results are robust within the neighborhood of our chosen $\lambda$.\\

\begin{figure}
\centering
\includegraphics[width=\textwidth]{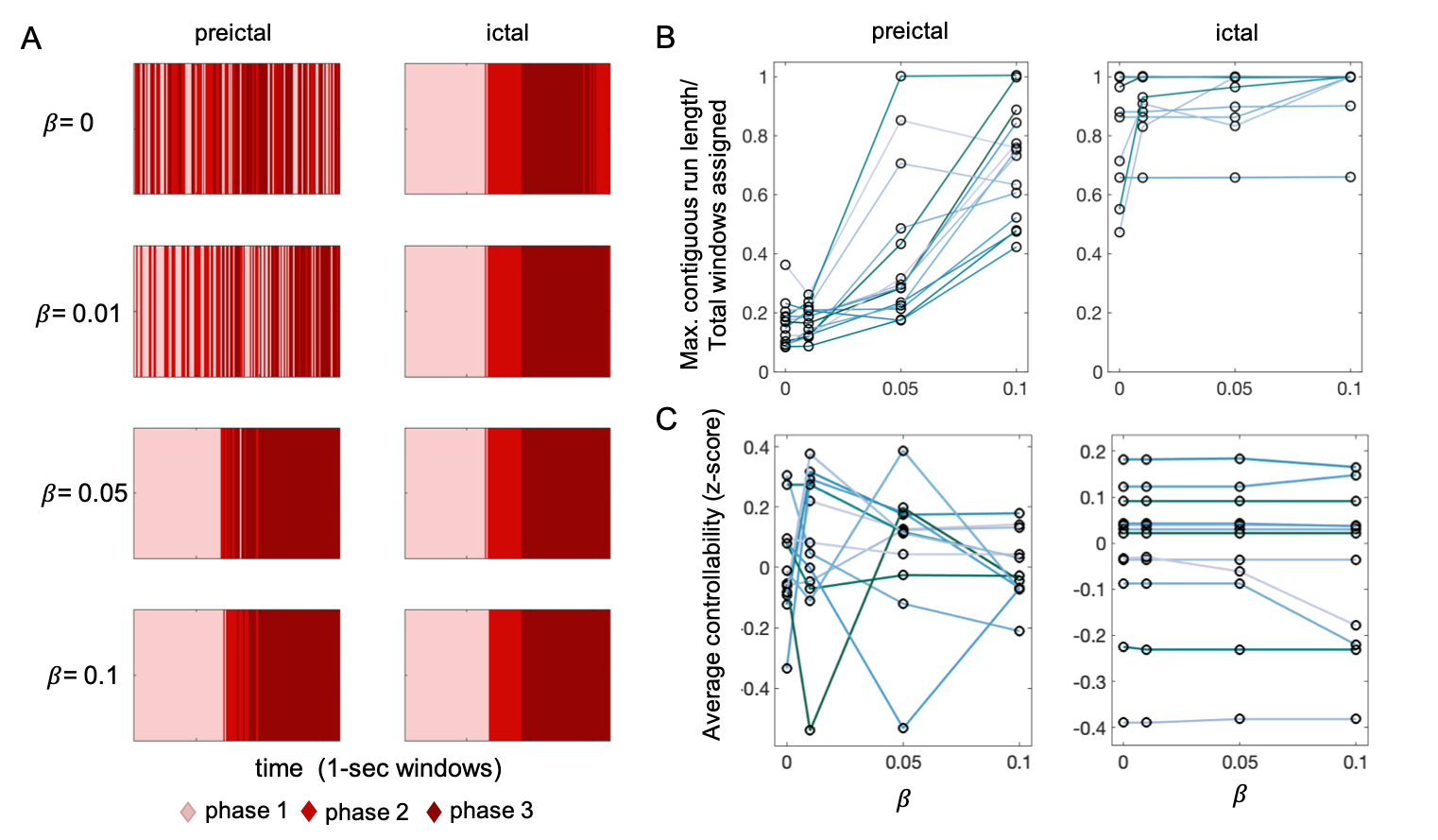}
\caption{\textbf{Sensitivity to phase contiguity parameter $\beta$.} (A) Temporal phase assignment with increasing $\beta$. Here we show the phase assignment to time windows in preictal and ictal periods of a sample seizure in patient “Study038”. At a weighting of $\beta$=0.01, the three seizure phases were temporally contiguous, while phases remained temporally interleaved in the preictal period. (B) Number of time windows in a temporally contiguous phase assignment normalized by total number of assigned time windows in the same phase, as a function of $\beta$ in preictal and ictal periods. A ratio of greater than 0.5 indicates that the majority of time windows in a phase are assigned to a single temporal cluster. Each line represents a single patient in the robustness subset $(n=14)$, different line colors are used for ease of visualization. (C) Change in phase 1 network average controllability values with increasing $\beta$. Each line represents a single patient in the robustness dataset. The difference between ictal average control values for $\beta$=0 and $\beta$=0.01 was not significant (Wilcoxon signed rank test $W=0,p=1.0$).}
\label{fig:beta}
\end{figure}

\noindent
\emph{Community detection contiguity parameter, $\beta$:} We performed community detection on similarity matrices built by correlating EC networks to determine transition points from seizure onset, propagation, and termination phases in a data-driven manner (see the later subsection "Modularity maximization of similarity matrices" in the Methods section of this Supplement). Increasing the parameter $\beta$ ensures that each of the three communities consistently represent seizure onset, propagation, and termination phases across patients by encouraging the discovered communities to be contiguous in time. We chose $\beta$=0.01 for our main results as this value ensured that at least 50\% of time windows assigned to a similarity community were temporally contiguous, yielding meaningful phases in the ictal period. Additionally at $\beta=0.01$ we found temporal interleaving of phase assignments in the preictal period, whereas greater values introduced an artificial temporal phase structure into the preictal period (Fig. \ref{fig:beta}A, B). Even as we tuned our distance weighting parameter $\beta$, any differences in phase assignment resulted in insignificant differences in controllability metrics across the 14 seizures of the robustness dataset (Fig \ref{fig:beta}C). For each metric, we used the Wilcoxon signed rank test to make a pairwise comparison between the metric values of phases assigned when $\beta$=0 and phases assigned when $\beta$=0.01. We found no difference in controllability values for any metric or phase $(n=14, W \leq 6, p > 0.25)$, indicating that our moderate contiguity adjustment does not change the main results. \\

\begin{figure}
\centering
\includegraphics[width=0.8\textwidth]{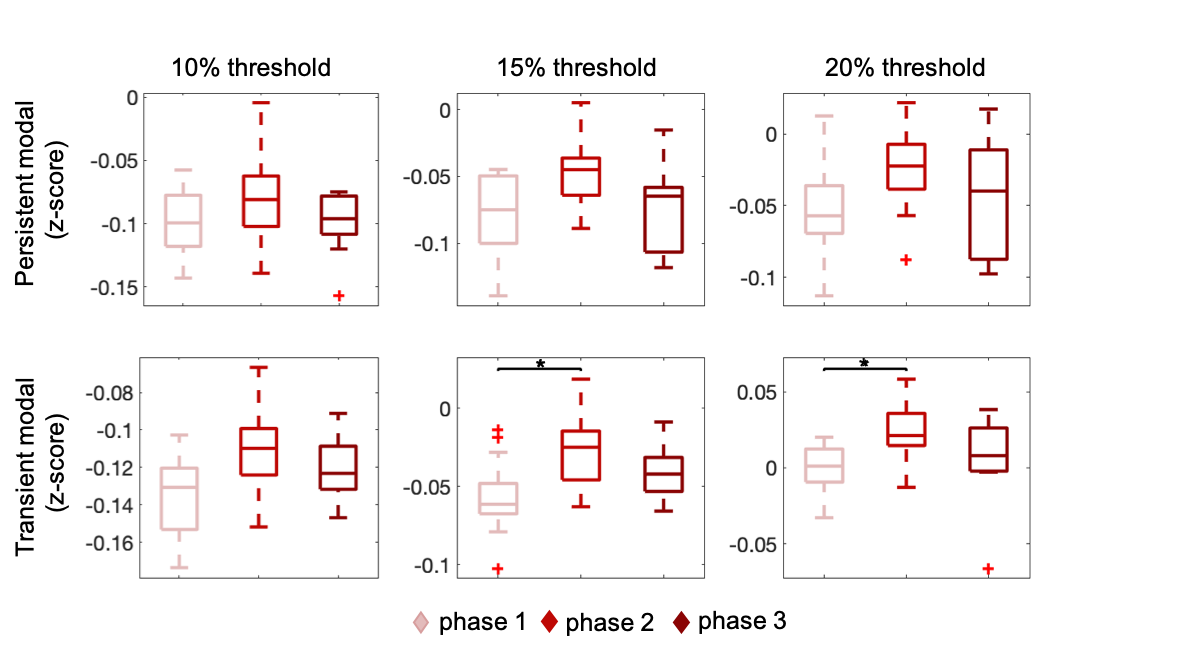}
\caption{\textbf{Variation of persistent and transient modal controllability threshold.} Persistent modal controllability and transient modal controllability values are obtained at a threshold that determines the number of highest (lowest) eigenvalues included in the calculation. We calculated group-level results for both metrics at thresholds of 10\%, 15\%, and 20\%. Boxplots indicate the 75\% confidence interval (box), median (solid line), 95\% confidence interval (whiskers), and outliers (stars). Starred bars indicate significant metric differences between phases using the $t$-test at the $p<0.017$ level. The difference between seizure onset and propagation was significant at the 15\% (Friedman's ANOVA $\chi^2(2, 14)=13.00, p<0.002$) and 20\% ($\chi^2(2, 14)=9.57, p<0.008$) thresholds of transient modal controllability. }
\label{fig:modal_thresh}
\end{figure}

\noindent
\emph{Persistent and transient modal controllability percentage:} In our main work we used the metrics of persistent modal controllability and transient modal controllability to further investigate the influence of a control node on the slow, sustaining modes of the system, or the fast, attenuating modes. We selected the top and bottom 15\% of the eigenvalues in a representative phase network to calculate persistent and transient modal controllability, respectively. Our threshold was chosen in an effort to capture variance in controllability values among nodes while maintaining a focus on the extremes of the eigenvalue distribution. To ensure that our results were robust to our choice of threshold, we reproduced our main group-level results at thresholds of 10\% and 20\%. Increasing (decreasing) the threshold increased (decreased) the persistent and transient modal metric values due to summing over a greater (fewer) number of eigenvalues as expected \cite{Karrer2019}. However we found that the relative group-level phase trends were maintained, with the propagation phase at a higher value for both metrics at each threshold (Fig \ref{fig:modal_thresh}). Even within the smaller robustness dataset, a significant difference between propagation and termination phases was found for transient modal controllability at a threshold of 15\% ($\chi^2(2, 14)=13.00, p<0.002$) and 20\% ($\chi^2(2, 14)=9.57, p<0.008$). This sensitivity analysis indicates that our main results are robust to choice of eigenmode threshold.\\

\begin{figure}
\centering
\includegraphics[width=0.8\textwidth]{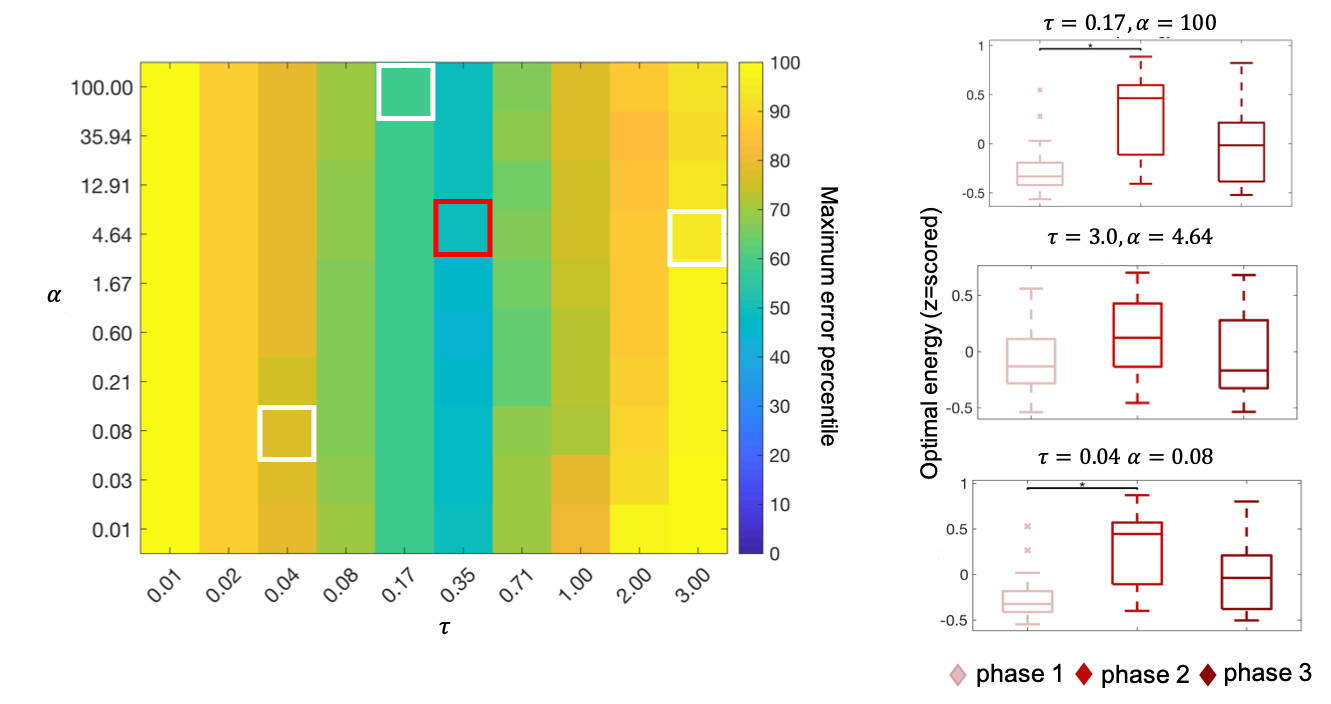}
\caption{\textbf{Maximum error percentile across all seizures and phases.} We used error magnitude to guide our parameter choices for computing optimal control energy and found that for each seizure phase across all seizures, the minimum error averaged across nodes occurred with parameter values in the neighborhood of $\tau \in [0.01, 3.00]$ and $\alpha \in [0.01, 100]$. \emph{Left}: We selected 10 values within each parameter interval and quantified the error percentile resulting from each parameter combination in a given seizure state. In our main work, we chose parameters of $\tau$=0.35 and $\alpha$= 4.46, because the computational error associated with optimal control energy estimation was within the smallest 46\% of all possible error values for every seizure phase ($Mdn=67.5, Q3-Q1=104$), and also represented an upper bound on error for any parameter combination for any seizure and phase. \emph{Right}: Results of group-level energy analysis are shown for three parameter combinations indicated in the left-hand plot. Boxplots indicate the 75\% confidence interval (box), median (solid line), 95\% confidence interval (whiskers), and outliers (stars). Starred bars indicate significant metric differences between phases using the $t$-test at the $p<0.017$ level. The parameter combination that did not demonstrate a significant change from onset to propagation phases was also associated with the largest computation error out of all test points ($Mdn=98.8, Q3-Q1=144.14$).
}
\label{fig:energy}
\end{figure}

\noindent
\emph{Optimal control energy time horizon $\tau$ and trade-off parameter $\alpha$:} We obtained estimates for the optimal control energy required to transition from brain activity during a seizure phase $x_0$ to a seizure-free state $x_\tau$. In our optimal control model, the control horizon $\tau$ specifies the time allotted to traverse the optimal trajectory. The distance-energy minimization tradeoff modifier, $\alpha$, can be tuned to prioritize a trajectory that minimizes the distance between $x_0$ and $x_\tau$ at each timestep, or to prioritize input energy minimization. We selected parameters of $\tau$ = 0.35 and $\alpha$= 4.46 for our main results, as this combination balanced estimation error across all seizures and phases (error $Mdn=67.5, Q3-Q1=104$), and a desire for a short trajectory (see the later subsection "Selection of the time horizon $\tau$ and parameter $\alpha$" in Methods section of this Supplement). We reproduced our group-level optimal control energy results for three additional parameter combinations across all patients ($n=34$) to ensure that our results were robust to our parameter choices. Figure \ref{fig:energy} displays group-level control energy results for three parameter pairs $(\tau,\alpha)=\{(3,4.64),(.04,.08),(0.17,100)\}$. The median and interquartile range of computational estimation error across seizures and phases for each combination were all greater than the error associated with our selected parameters, at $(98.7, 144.14), (95.16, 140.45)$, and $(81.23, 124.17)$, respectively. Despite the increased estimation error, a significant increase from seizure onset to propagation was found for parameters (0.04, 0.08), $\chi^2(2,34)=21.23,p=2.4\times 10^{-5})$, and for parameters (0.17,110), $\chi^2(2,34) =17.17,p<1.8\times 10^{-4})$. This increase was followed by a subsequent decrease from propagation to termination phases, a result consistent with our main findings. 

\section*{Supplementary Methods}
\subsection*{Estimation of effective connectivity}
In this study, we were motivated to use an effective connectivity measure that was related to the notion of a partial correlation. In its common form, a partial correlation matrix describes all pairwise partial correlations between nodes in a network, and is the standardized form of the inverse covariance matrix, $\boldsymbol{\boldsymbol{\Sigma}}^{-1}$, where $\boldsymbol{\boldsymbol{\Sigma}}_{i,j}^{-1}=0$ implies conditional independence between nodes $i$ and $j$ given the activity of all other nodes in the network. Thus, partial correlation can model unique interactions between nodes and can be indicative of causal links in the network \cite{Epskamp2018}.

Unfortunately, the number of ECoG recording channels is smaller than the number of samples in a time window, which means that the inverse covariance matrix cannot be directly computed, and must instead be estimated. The Graphical Least Absolute Shrinkage and Selection Operator (GLASSO)\cite{Banerjee2007,Friedman2008} method for regularized inverse covariance estimation is widely implemented due to its efficiency and simplicity \cite{colclough2018,Rosa2015,Smith2011,Wang2018}, and is what we implement in our work. The method uses a Gaussian graphical modeling approach to compute a regularized inverse covariance estimate $\boldsymbol{\Omega} \approx \boldsymbol{\boldsymbol{\Sigma}}^{-1}$ from a set of variables assumed to have a Gaussian distribution with mean $\mu$ and covariance $\boldsymbol{\boldsymbol{\Sigma}}$. GLASSO maximizes the penalized log-likelihood
\begin{equation}
	\text{log det} \boldsymbol{\Omega} - \text{tr}(\boldsymbol{\Omega}\boldsymbol{\Sigma})-\rho ||\boldsymbol{\Omega}||_1
\end{equation}
over all nonnegative definite matrices $\boldsymbol{\Omega}$ from the sample covariance matrix $\boldsymbol{\boldsymbol{\Sigma}}$. Here $\text{log det}$ is the logarithm of the determinant, $\text{tr}$ is the trace of the matrix, $||.||_1$ is the matrix L1-norm (sum of the absolute values of all entries), and $\rho$ controls the sparsity of the $\boldsymbol{\Omega}$ estimate. We choose the value of the sparsity parameter $\rho$ using extended Bayesian Information Criterion (EBIC) for model selection. EBIC requires tuning of a single hyperparameter, $\lambda$, to select for simpler models with fewer edges or more exploratory models with a greater chance of false discovery. In our main work we choose a conservative value of $\lambda= 0.5$, at the end of the standard parameter range that prefers a low false discovery rate at the cost of missing some true edges \cite{Foygel2010}. 

The R package bootnet \cite{Epskamp2018} for high-dimensional undirected graph estimation was used to compute effective connectivity matrices from the channel timeseries data \cite{ZhaoHaoming2019}. Through the package we first applied a non-paranormal transformation to the data to relax the assumption of normality \cite{Liu2009}. Then, we implemented the GLASSO function to compute the regularized inverse covariance matrix for 100 values of $\rho$, and used EBIC to select the optimal sparsity tuning parameter between 0.01 and 0.1 that would find an accurate regularized inverse covariance estimate. Finally, the regularized inverse covariance matrices, $\omega$, were standardized to generate effective connectivity matrices, where the effective connectivity (EC) between variables $y_i$ and $y_j$ conditioned on all other variables $y_{-(i,j)}$ is given by 
\begin{equation}
	EC(y_i,y_j | y_{-(i,j)})=-\frac{\boldsymbol{\Omega}_{i,j}}{\sqrt{\boldsymbol{\Omega}_{i,i}} \sqrt{\boldsymbol{\Omega}_{j,j}}} .
\end{equation}
The EC networks resulting from our estimation pipeline provided the foundation for the metric analysis and optimal control model explored in this study.

\subsection*{Modularity maximization of similarity matrices} 
\label{sec:modularitySection}
We took a data-driven community detection approach to determine the three groups of effective connectivity networks that shared a similar structure within a seizure. As we described in the main text, for each seizure of variable length $T$ seconds recorded with a variable $N$ channels, we constructed $T$ effective connectivity matrices of size $N \times N$. Next, we calculated a $T \times T$ network similarity matrix, $\boldsymbol{S}$, from the chronologically ordered EC networks, where the $ij-th$ element represents the Pearson correlation coefficient between the upper triangle of the $i-th$ EC network and the upper triangle of the $j-th$ EC network.

Next, we weighted each cell in the similarity matrix by a squared distance weighting. The weighting of the $ij-th$ element is given by
\begin{equation}
w_{i,j}=(1-\beta \frac{|i-j|}{N})^2,
\end{equation}
where $\beta$ tunes the magnitude of $w_{i,j}$. This distance weighting ensured that each of the three communities consistently represented seizure onset, propagation, and termination phases across patients by encouraging the discovered communities to be contiguous in time.

To perform community detection on these weighted similarity matrices, we used a Louvain-like locally greedy algorithm to maximize the following modularity quality function: 
\begin{equation}
	Q=\sum_{i,j}(\boldsymbol{S}_{i,j}-\gamma \boldsymbol{P}_{i,j} )\delta(g_i,g_j),
\end{equation}
where $\boldsymbol{S}$ is the weighted network similarity matrix, node $i$ is assigned to community $g_i$, node $j$ is assigned to community $g_j$, and $\delta$ is the Kronecker delta function where $(g_i,g_j)=1$ if $g_i=g_j$, and 0 otherwise \cite{Blondel2008}. The structural resolution parameter ${\gamma}$ tunes the relative importance of the null model $\boldsymbol{P}$. Here we use the Newman-Girvan null model of the form $\boldsymbol{P}_{i,j}=\frac{k_i k_j}{2m}$, where $k_i=\boldsymbol{\Sigma}_j \boldsymbol{S}_{i,j}$ , and $m=\frac{1}{2}\boldsymbol{\Sigma}_{i,j}\boldsymbol{S}_{i,j}$ \cite{Girvan2002}. Greater values of $\gamma$ result in more discovered communities and lower maximized modularity, $Q_{max}$ \cite{Reichardt2006}. We initially let $\gamma$ take values in the range of 0.8 to 1.0 in increments of 0.05, and either expanded the range or increased the range resolution to find a value of $\gamma$ that resulted in the discovery of three communities. Finally, whenever we observed a given community to not be contiguous in time, we selected the longest run of adjacent networks assigned to the community to represent the seizure phase. On average only 3.2\% ($n=34, Mdn=0\%, Q3-Q1=2.7\%)$ of the time windows in a given seizure were not assigned to the longest contiguous run of any phase. 

\subsection*{Controllability metrics}
The main objective in this study was to leverage the effective connectivity framework to indicate seizure phases and cortical locations that are optimal for controlling seizure dynamics with external stimuli. We used techniques from linear control theory to understand how the underlying network topology would respond to external input energy. A linear system is controllable if it can be driven from an initial state $x_0$ to a desired target state $x_\tau$ within a finite time period. We calculated a number of controllability metrics that quantify different aspects of system control and are derived from the spectral properties of the effective connectivity networks and controllability Gramian. We formally define each of them in turn below.

The notion of \emph{average controllability} that we use is a measure of the energy contained in the impulse response of the system when energy is injected through a set of control nodes $\boldsymbol{B}_K$. The metric is inversely proportional to the average input energy required for nodes in the control set to push the system into all possible states with unit norm \cite{Gu2015}. Average controllability is calculated as Trace($\boldsymbol{W}_K$), where $\boldsymbol{W}_K$ is the controllability Gramian of a system with $K$ control points, which in turn is defined as
\begin{equation}
	\boldsymbol{W}_K= \int_{0}^{\infty} {e^{\boldsymbol{A}t} \boldsymbol{B}_K \boldsymbol{B}_K^T e^{\boldsymbol{A}^Tt}  dt}.
\end{equation}

In the context of neural stimulation with the brain represented by a linear system, the modes of the system will reflect how the brain attenuates or sustains external perturbations. Modal controllability describes how much influence a single control node has on driving all modes of the system. This statistic is calculated as the scaled measure of controllability from all nodes by a single node $i$, which can be defined as
\begin{equation}
	\phi_i=\sum_{j}(1-(e^{\lambda_j})^2 ) v_{ij}^2,
\end{equation}
where $\boldsymbol{V}=[v_{ij}]$ is the eigenvector matrix of the adjacency matrix $\boldsymbol{A}$, $\lambda_j$ is the $j$\-th eigenvalue of $\boldsymbol{A}$, and $e^{\lambda_j}$ converts the continuous system eigenvalues to a discrete representation \cite{Pasqualetti2014}. 

We used the additional metrics of persistent modal controllability and transient modal controllability to further investigate the influence of a control node on the slow, sustaining modes of the system, or the fast, attenuating modes\cite{Karrer2019}. We separated modal controllability into persistent modal controllability and transient modal controllability by restricting the calculation in the above equation (8) to only those eigenvectors corresponding to a set of the 15\% largest or smallest eigenvalues, respectively \cite{Tang2019}. Persistent controllability will be high for control nodes that cause a large response in the slow modes of the system, and transient controllability will be high for control nodes that cause a large response in the fast modes of the system. 

In each network, we calculated average, modal, persistent modal, and transient modal controllability using a relaxed control set $\boldsymbol{B}_K$ \cite{Stiso2018}, which considers one node $n_c$ as a primary control point such that $\boldsymbol{B}_{n_c,n_c}=1$, with all other nodes set to a small relative control contribution of $10^{-7}$. Controllability metrics were calculated considering each node as a control node, respectively, to obtain an $N \times 1$ controllability vector for every one-second time window throughout a seizure. We then averaged the controllability vectors within a given seizure phase across time to obtain the representative $N \times 1$ controllability vector for that phase.

\subsection*{Optimal control energy}
We used the concept of optimal control energy to understand how stimulation should modulate the brain over time to move it from a seizure phase to a seizure-free state. Optimal control energy describes the energetic input required to move a complex system from an initial state $x_0$ to a desired target state $x_\tau$ along an optimal trajectory within a finite time horizon $\tau$. In our work, we modeled the optimal control energy required to drive the brain state during a particular seizure phase into a seizure-free brain state sampled from the pre-ictal period. Following previous work, we represented the initial brain state $x_0$ as an $N \times 1$ vector of band power values collected over consecutive one-second time windows for each channel and averaged across all time windows in a given seizure phase \cite{Stiso2018}. We defined $x_\tau$ to be the $N \times 1$ band power vector averaged across all pre-ictal time windows. In the main manuscript, we specifically consider brain state in the high-$\gamma$ frequency band (95-105 Hz). This measure of band power has been used extensively for epilepsy detection as it is sensitive to variation of both signal frequency and amplitude \cite{Weiss2013, Baldassano2017,Kohling2000, Weiss2015}. We used it here to approximate initial and final brain state as it provided a convenient mechanism to relate relative activity among nodes within and across seizure phases.

Next, we defined the optimal trajectory as the path that minimized both control energy and the distance between an intermediary state $x(t)$ and a final state $x_\tau$. The optimal control function $u_K^*(t)$ solves this linear-quadratic optimal control problem where the objective is to minimize the following cost function over $u$:
\begin{equation}
	J=x_\tau^T x_\tau+\int_{0}^{\tau} (x(t)-x_\tau )^T (x(t)-x_\tau )+\alpha u(t)^T u(t) dt)
\end{equation}
subject to 
$\dot{x}(t)=\boldsymbol{A}x(t)+\boldsymbol{B}u(t), t \in [0,\tau], x(0)=x_0, x(\tau)=x_\tau$.

Here the parameter $\alpha$ balances the importance of minimizing total energy against following the most direct path. As was done for the metric calculations, we treated each node as a control node, respectively, while creating a relaxed control input $\boldsymbol{B}$ matrix by setting the control contributions of all other nodes to $10^{-7}$ along the matrix diagonal \cite{Stiso2018}. We obtained an estimate of $u_K^*(t)$ using the Pontryagin Minimum Principle from the MATLAB script optim\_fun.m at https://github.com/jastiso. Finally, we used $u_K^*(t)$ to compute optimal control energy as
\begin{equation}
	E_{opt}=\int_{0}^{\tau} ||u_K^*(t)||^2 dt.
\end{equation}
\noindent Thus, we obtained an $N \times 1$ vector of optimal control energy values for each seizure phase, representing the energy required to move the system to a seizure-free state by driving a single node with the optimal control function. 

\subsection*{Selection of the time horizon $\tau$ and parameter $\alpha$}
\label{sec:energyParamSelection}
Both the control horizon, $\tau$, and the distance-energy minimization tradeoff modifier, $\alpha$, are free parameters in our optimal energy model. We wished to select a value for $\tau$ that allowed the system to expeditiously converge to the target state, as this would reflect a fast transition to a seizure-free activity level. Solving for the optimal control function can become computationally expensive for large values of $\tau$, and the estimation of matrix exponentials that is involved when using the Pontryagin Minimum Principle here can lead to estimation errors. In the case where $\tau$ is small, the computational error also reflects the inability of the system to converge to $x_\tau$. We used error magnitude to guide our parameter choices and found that for each seizure phase across all seizures, the minimum error averaged across nodes occurred with parameter values in the neighborhood of $\tau \in [0.006,3.0]$ and $\alpha \in [0.001,100]$. We selected 10 values within each parameter interval and quantified the error percentile resulting from each parameter combination in a given seizure state. We chose parameters of $\tau$ = 0.35 and $\alpha$= 4.46 because the associated computational error was within the smallest 46\% of all possible error values for every seizure phase, and because it represented the minimum achievable bound on error for any parameter combination.

\begin{figure}
\centering
\includegraphics[width=0.5\textwidth]{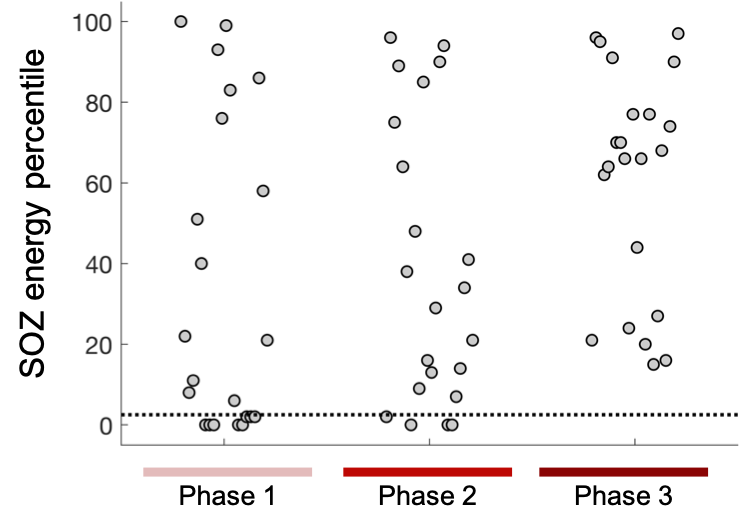}
\caption{\textbf{Optimal control energy using SOZ nodes as the control set.} We measured the total energy required to drive the brain state from a given seizure phase $x_0$ to a final seizure-free state $x_{\tau}$ resulting from simultaneous control of all SOZ points. We then repeated this measurement in 5000 permutations of the SOZ labels among the non-SOZ electrodes. We found that in half of the patients with localized seizures, control using the SOZ nodes required significantly less energy than control from other subsets of nodes in the seizure onset phase $(N=5000, p<0.025)$. The distribution of SOZ energy percentile with respect to a null distribution for the 22 seizures with marked SOZs is shown for each phase. The dashed line is located at the 97th lowest percentile.}
\label{fig:sozEnergy}
\end{figure}

\begin{table}\centering
\caption{Patient Information}

\begin{tabular}{lrrrrrrr}
Patient ID & Sex & Age at Surgery & Onset Location & Seizure Type & N Seizures & N SOZ Nodes & N Channels \\
\midrule
	HUP64 &	M	& 20	&	Left Frontal	&	CPS+GTC	&	1	&	4	&	64\\
	HUP65 &	M	& 37	&	Right Temporal	&	CPS+GTC	&	3	&	12	&	63\\
	HUP68 &	F	& 26	&	Right Temporal	&	CP+GTC	&	4	&	5-12	&	63\\
	HUP74 &	F	& 25	&	 Middle Temporal	&	CPS	&	1	&	5	&	63\\
	HUP80 &	F	& 27	&	Bilateral Left	&	CP+GTC	&	4	&	8-12	&	63\\
	HUP86 &	F	& 25	&	Left Temporal	&	CP+GTC	&	2	&	22	&	60\\
	HUP099 &	F	& 20	&	Right Temporal	&	CPS+GTC	&	1	&	--	&	57\\
	HUP106 &	F	& 45	&	Left Temporal	&	CPS+GTC	&	3	&	--	&	64\\
	HUP107 &	M	& 36	&	Right Temporal	&	CPS	&	1	&	--	&	64\\
	HUP119 &	F	& 57	&	--	&	CPS	&	2	&	--	&	56	\\
	Study010 &	F	& 13	&	Left Frontal	&	GTC	&	2	&	10	&	46\\
	Study023 &	M	& 16	&	Left Occipital	&	GTC	&	2	&	2	&	63\\
	Study026 &	M	& 9	&	Left Frontal	&	CPS+GTC	&	5	&	0-1	&	56\\
	Study038 &	M	& --	&	Left Frontal	&	CPS+GTC	&	3	&	3-6	&	48\\
\bottomrule
\end{tabular}
\end{table}

\FloatBarrier

\bibliography{supplement_bib}